\documentclass[aps,twocolumn,a4paper,floatfix,superscriptaddress,pra,longbibliography]{revtex4-2}
\usepackage[toc,page]{appendix}
\usepackage{braket}
\usepackage{epsfig,graphicx,times}
\usepackage{amstext}
\usepackage{amsmath}
\usepackage{amssymb}
\usepackage{mathrsfs}
\usepackage{dcolumn}
\usepackage{bm}
\usepackage[tight]{subfigure}
\usepackage[colorlinks,linkcolor=blue,anchorcolor=blue,citecolor=blue,urlcolor=blue]{hyperref}
\usepackage{color}
\usepackage{siunitx}
\usepackage{appendix}
\usepackage{placeins}
\usepackage{textcomp}
\usepackage{bbold}
\usepackage{float}
\usepackage{verbatim}
\usepackage{minitoc}
\usepackage[dvipsnames]{xcolor}

\usepackage{soul}
\usepackage[framemethod=tikz]{mdframed}
\usepackage{lipsum}
\usepackage{epstopdf}

\begin{document}

\title{Phase-selective tripartite entanglement and asymmetric Einstein-Podolsky-Rosen steering in squeezed optomechanics}

\author{Ya-Feng Jiao}
\affiliation{School of Electronics and Information, Zhengzhou University of Light Industry, Zhengzhou 450001, China}
\affiliation{Academy for Quantum Science and Technology, Zhengzhou University of Light Industry, Zhengzhou 450001, China}

\author{Jie Wang}
\affiliation{Key Laboratory of Low-Dimensional Quantum Structures and Quantum Control of Ministry of Education, \\
Department of Physics and Synergetic Innovation Center for Quantum Effects and Applications, \\ Hunan Normal University, Changsha 410081, China}

\author{Dong-Yang Wang}
\affiliation{School of Physics, Zhengzhou University, Zhengzhou 450001, China}

\author{Lei Tang}
\affiliation{College of Physics and Electronic Engineering, Institute of Solid State Physics, Sichuan Normal University, Chengdu 610101, China}

\author{Yan Wang}
\affiliation{School of Electronics and Information, Zhengzhou University of Light Industry, Zhengzhou 450001, China}
\affiliation{Academy for Quantum Science and Technology, Zhengzhou University of Light Industry, Zhengzhou 450001, China}

\author{Yun-Lan Zuo}
\affiliation{School of Physics and Chemistry, Hunan First Normal University, Changsha 410205, China}

\author{Wan-Su Bao}\email{bws@qiclab.cn}
\affiliation{Henan Key Laboratory of Quantum Information and Cryptography, IEU, Zhengzhou 450001, China}

\author{Le-Man Kuang}\email{lmkuang@hunnu.edu.cn}
\affiliation{Academy for Quantum Science and Technology, Zhengzhou University of Light Industry, Zhengzhou 450001, China}
\affiliation{Key Laboratory of Low-Dimensional Quantum Structures and Quantum Control of Ministry of Education, \\
Department of Physics and Synergetic Innovation Center for Quantum Effects and Applications, \\ Hunan Normal University, Changsha 410081, China}

\author{Hui Jing}\email{jinghui73@gmail.com}
\affiliation{Academy for Quantum Science and Technology, Zhengzhou University of Light Industry, Zhengzhou 450001, China}
\affiliation{Key Laboratory of Low-Dimensional Quantum Structures and Quantum Control of Ministry of Education, \\
Department of Physics and Synergetic Innovation Center for Quantum Effects and Applications, \\ Hunan Normal University, Changsha 410081, China}

\date{\today}

\begin{abstract}
The generation and manipulation of multipartite entanglement and EPR steering in macroscopic systems not only play a fundamental role in exploring the nature of quantum mechanics, but are also at the core of current developments of various nascent quantum technologies. Here we report a theoretical method using squeezing-phase-controlled quantum noise flows to selectively generate and manipulate quantum entanglement and asymmetric EPR steering in a nonlinear $\chi^{(2)}$ whispering-gallery-mode (WGM) optomechanical resonator. We show that by pumping the $\chi^{(2)}$ nonlinear medium with two-photon optical fields and broadband squeezed lights, a pair of counterpropagating squeezed optical modes could be introduced to the WGM resonator, each coupled with an independent squeezed vacuum reservoir. This configuration could enable squeezing-phase-controlled light-reservoir interaction for each squeezed optical mode, providing a flexible tool for tailoring asymmetric optical noise flows in the counterpropagating modes. Based on this unique feature, it is found that with the injection of asymmetric noise flows, the generation of various types of bipartite and tripartite entanglement become phase-dependent and thus they can be produced in an asymmetric way. More excitingly, it is also found that by further properly adjusting the squeezing parameters, the overall asymmetry of EPR steering can also be stepwise driven from no-way regime, one-way regime to two-way regime. These findings, holding promise for preparing rich types of entangled quantum resources with asymmetric features, may have potential applications in the area of secure quantum information processing such as quantum secure direct communication and one-way quantum computing.
\end{abstract}

\maketitle

\section{Introduction}

Entanglement, allowing perfectly correlated positions and momenta for two spatially separated particles, has long been intriguing in quantum physics and enables numerous advanced quantum information protocols spanning from quantum networking to quantum sensing~\cite{Horodecki2009RMP}. The concept of entanglement was originally addressed by Schr\"{o}dinger~\cite{Schrodinger1935MPCPS} in his response to the issue of the ``spooky action-at-a-distance" predicted by Einstein, Podolsky, and Rosen (EPR) in their famous paradox~\cite{Einstein1935PR}, where he also coined a term that came to be known as the EPR steering~\cite{Reid1989PRA,Ou1992PRL}. From the perspective of violations of local-hidden-state models, Wiseman \textit{et al.} has formalized an operational benchmark for EPR steering~\cite{Wiseman2007PRL}, by which they further proved that under the hierarchy of quantum nonlocality, EPR steering is a strict subset of entanglement and a strict superset of Bell nonlocality. An appealing feature of EPR steering is that it describes how local measurements on one part of the system can steer (alter) the state of the other part at a different location. This defining characteristics, not held by the other two types of quantum nonlocality, reveals the intrinsic asymmetry of EPR steering and offers an insight into directional nonlocality~\cite{He2015PRL}, which plays an indispensable role in enabling quantum techniques using untrusted devices~\cite{Xiang2022PRXQ}, such as secure quantum key distribution~\cite{Branciard2012PRA,Nathan2016Optica}, randomness certification~\cite{Li2024PRL}, and no-cloning quantum teleportation~\cite{He2015PRL2,Chiu2016NQI}. In the past few decades, after a series of rigorous mathematical characterizations of quantum nonlocality~\cite{Simon2000PRL,Adesso2004pra,Adesso2007JPA,Kogias2015PRL}, a great deal of progresses have been made experimentally to prepare entangled or steerable states of microscopic and macroscopic particles, involving platforms based on photons~\cite{Wang2016PRL}, ions~\cite{Stute2012Nature}, atoms~\cite{van2022Nature}, superconducting circuits~\cite{Kurpiers2018Nature}, and cavity optomechanical (COM) devices~\cite{Palomaki2013Science,Kotler2021Science,Mercier2021Science}. However, in terms of generation and manipulation of macroscopic entanglement, it is still challenging to avoid the decoherence effect induced by device imperfection. Very recently, to overcome this obstacle and achieve entangled state with high fidelity, a large number of theoretical proposals have been raised, which relies on synthetic gauge field~\cite{Lai2022PRL,Liu2023SCMPA}, reservoir engineering~\cite{Wang2013PRL,Yang2015PRA}, dark-mode or feedback control~\cite{Lai2022PRR,Huang2022PRA2,Li2017PRA}, photon counting~\cite{Ho2018PRL}, dynamical modulation~\cite{Wang2016PRA,Yang2024OE}, injection of quantum squeezing~\cite{Jiao2024LPR, Wu2024OE}, and optical nonreciprocity~\cite{Jiao2020PRL,Jiao2022PRAPP,Jiao2025FR}.

On the other hand, squeezed light, characterized by reduced quantum noise in selected quadratures, has emerged as a vital resource in modern quantum science~\cite{Scully1997book}. Squeezed light has been widely utilized to cool mechanical motion~\cite{Clark2017Nature}, enhance light-matter interaction~\cite{Qin2018PRL,Li2020PRL,TangL2022PRL}, and advance quantum information processing~\cite{Andersen2010LPR,Flamini2018RPP} across a broad range of quantum platforms, including optomechanical devices~\cite{Lau2020PRL}, cavity QED systems~\cite{Murch2013Nature}, and trapped ions~\cite{Ge2019PRL}. In particular, the peculiar feature of quadrature noise reduction make squeezed light an indispensable resource for improving the sensitivity of quantum measurements, leading to practical applications in gravitational-wave detection~\cite{Aasi2013NP}, nanoscale sensing~\cite{Zhang2024SCPMA}, and nondemolition qubit readout~\cite{Didier2015PRL}. In practice, squeezed light is typically introduced to quantum systems either by external injection or by internal generation via $\chi^{(2)}$ nonlinearity. However, for current experimental techniques, both approaches face significant limitations: external injection suffers from inevitable transmission and coupling losses that render squeezed states fragile, while intracavity generation is fundamentally constrained by the 3 dB limit~\cite{Collett1984PRA}, which prevents quantum noise of a cavity field from being reduced below half of the zero-point fluctuations. Recently, L\"{u} \textit{et al.} proposed a novel squeezed-light scheme that combines external injection and intracavity generation~\cite{Lu2015PRL}, effectively addressing the limitations of the conventional methods and enabling single-photon quantum processes in the weak-coupling regime of an optomechanical system. Building on this approach, subsequent studies have further explored the potential of this hybrid squeezing strategy in various quantum applications, such as surpassing the 3 dB limit of intracavity squeezing~\cite{Qin2022PRL}, enhancing qubit readout~\cite{Qin2024PRL}, and generating long-lived cat states~\cite{Chen2021PRL,Qin2021PRL}.

Inspired by these studies, we here investigate how to achieve selective generation of tripartite entanglement and asymmetric EPR steering through tuning squeezing phase in a WGM optomechanical resonator. The WGM resonator is made of $\chi^{(2)}$ nonlinear materials and supports two degenerate counterpropagating optical modes and two mechanical modes. Particularly, we show that by pumping the $\chi^{(2)}$ nonlinear medium with two-photon optical fields and broadband squeezed lights, a pair of counterpropagating squeezed optical modes could be introduced to the WGM resonator, each coupled with an independent squeezed vacuum reservoir. This configuration enables a squeezing-phase-controlled light-reservoir interaction for each squeezed optical mode and provides a flexible tool for manipulating the flow of optical input noises,which plays a key role in producing phase-selective entanglement. Particularly, it is found that when tailoring asymmetric flow of optical noises in different input directions, the generation of various types of bipartite and tripartite entanglement become phase-dependent and they can only be generated in an asymmetric way. More interestingly, by properly adjusting the squeezing phase-matched condition, we further show that the overall asymmetry of EPR steering can be stepwise driven from no-way regime, one-way regime to two-way regime. These generated asymmetric entangled states, with their rich variety, are applicable to a wide range of on-chip secure quantum information protocols~\cite{Branciard2012PRA,Nathan2016Optica,Li2024PRL,He2015PRL2,Chiu2016NQI}. Besides, our work, providing an efficient all-optical approach to engineer quantum states in an asymmetric way, can also be extended to explore the manipulation of other types of quantum effect, such as photon blockade~\cite{Huang2018PRL,Yang2023LPR,Wang2023OE} and quantum phase transition~\cite{Fruchart2021Nature,Zhu2024PRL}. As such, we believe that our proposal would not only be promising for serving in fundamental tests of quantum theories, but also can serve as key quantum resources for nascent quantum technologies like quantum secure communication~\cite{He2015PRL2,Chiu2016NQI} and one-way quantum computing~\cite{Raussendorf2001PRL,Walther2005Nature}.

\begin{figure*}[t]
\centering
\includegraphics[width=0.95\textwidth]{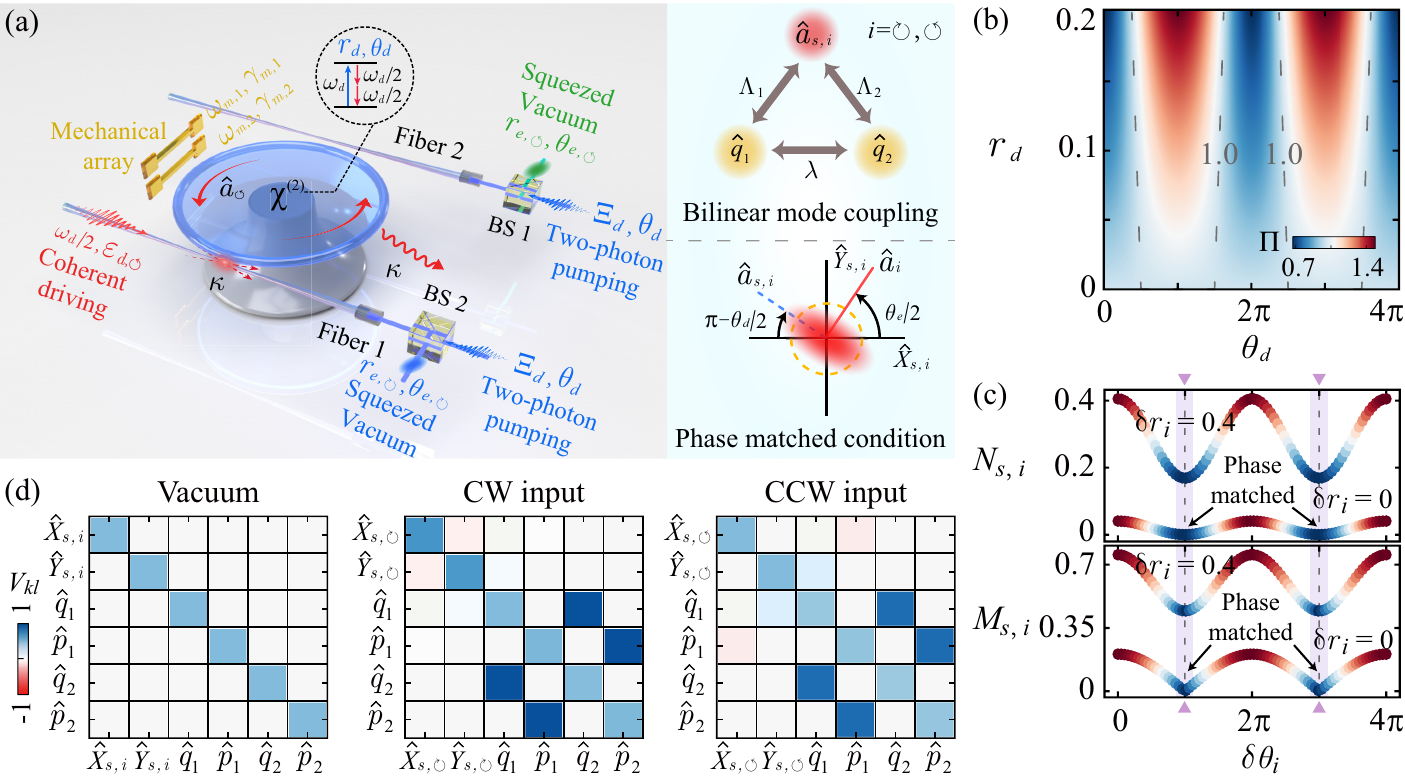}
\caption{\label{Fig1}(a) Schematic of a squeezed COM setup, showing a tapered-fibers-interfaced microdisk WGM resonator dispersively coupled to a string of nanomechanical oscillators. The WGM resonator is made of $\chi^{(2)}$ nonlinear materials, supporting two degenerate counterpropagating modes. The mechanical string offers two mechanical modes, which can be excited by placing the string in the evanescent field of the WGM resonator. When applying two-photon pumping fields along with broadband squeezed lights from the right fiber ports, two squeezed counterpropagating modes $\hat{a}_{s,\circlearrowright}$ and $\hat{a}_{s,\circlearrowleft}$ could be introduced in the WGM resonator, which are coupled with independent squeezed vacuum reservoirs. The coherent driving field is input from the left fiber port and is coupled into or out of the WGM resonator via evanescent coupling. (b) Density plot of the enhancement factor $\Pi$ as a function of squeezing strength $r_{d}$ and squeezing reference angle $\theta_{d}$. The effective COM interaction can be either enhanced ($\Pi>1$) or reduced ($\Pi<1$) depending on squeezing parameters. (c) Plot of the effective optical thermal noise $N_{s,i}$ and the effective two-photon correlation strength $M_{s,i}$ versus the phase difference $\delta\theta_{i}$ for different values of $\delta r_{i}$. Specifically, for the CCW input case, we consider the squeezed optical mode $\hat{a}_{s,\circlearrowleft}$ is phase-matched with its squeezed vacuum reservoir, leading to $N_{s,\circlearrowleft}=M_{s,\circlearrowleft}=0$. For the CW input case, it is phase-mismatched for $\hat{a}_{s,\circlearrowright}$, leading to $N_{s,\circlearrowright}\neq0$ and $M_{s,\circlearrowright}\neq0$. (d) The steady-state CM $V$ for opposite input cases. The parameters used here are given in the main text.}
\end{figure*}

This paper is structured as follows. In Sec.\,\ref{secII}, we introduce the theoretical model of the proposed COM system and obtain the effective Hamiltonian and master equation of this system, by which we calculate the system dynamics and evaluate the quantitative measures for entanglement and EPR steering. In Sec.\,\ref{secIII}, based on the numerical simulations, we analyze the method and mechanism for achieving phase-selective generation and manipulation of various types of entanglement and EPR steering. In Sec.\,\ref{secIV}, a brief summary is given. The detailed derivation process of the effective Hamiltonian, the associated master equation, and the system dynamics is given in the Appendix.

\section{Theoretical model and system dynamics}\label{secII}
In this paper, we propose how to use phase-controlled quantum noise flows to selectively generate and manipulate quantum entanglement and asymmetric EPR steering in a three-mode COM system with two mechanical modes. Here, as depicted in Fig.\,\ref{Fig1}(a), we consider a near-field COM system as a possible experimental setup~\cite{Anetsberger2009NP,Sbarra2022NL}, which consists of a tapered-fibers-interfaced microdisk WGM resonator dispersively coupled to a string of nanomechanical oscillators. The microdisk WGM resonator is assumed to be made up of $\chi^{(2)}$ nonlinear materials, e.g., lithium niobate or aluminum nitride, thereby supporting optical parametric down-conversion process. Due to the spatial symmetry of the microdisk WGM resonator, it supports two degenerate counterpropagating optical modes with resonance frequency $\omega_{c}$, i.e., the clockwise (CW) and counterclockwise (CCW) modes, respectively. For high-quality WGM resonators without backscattering losses, the CW and CCW modes could keep uncoupled. The two mechanical modes with fundamental frequency $\omega_{m,j}$ $(j=1,2)$ are offered by the nanomechanical oscillators in the mechanical string. As demonstrated in the experiment~\cite{Anetsberger2009NP}, by placing such mechanical string close to the tightly confined evanescent field of the microdisk WGM resonator, each oscillator can excite a mechanical vibration mode in the direction orthogonal to the microdisk surface, which is resulting from the evanescent field induced COM interaction. Furthermore, due to the high-finesse of optical and mechanical modes, as well as the mitigated losses and thermal effects during the coupling process, the observed COM interaction in this experiment is purely dispersive. In addition, given that the two mechanical modes are degenerate and interact with the same cavity mode, an effective phonon-hopping interaction could be generated between them, which is mediated by the cavity field induced hybridization effect~\cite{Shkarin2014PRL}. Note that such optomechanically induced hybridization of distinct mechanical modes has been experimentally demonstrated in nano-optomechanical structures~\cite{Lin2010NP}, WGM resonators~\cite{Pennetta2020PR,Zhang2012PRL}, and multimode circuit systems~\cite{Massel2012NC}.

As shown in Fig.\,\ref{Fig1}(a), the microdisk WGM resonator can be driven from two opposite input directions. The coherent driving field with frequency $\omega_{d}/2$ applied to port $1$ (port $2$) could be evanescently coupled into the CCW (CW) mode, which is referred to as the CCW (CW) input case. For the CCW input case, the Hamiltonian of the whole system in a frame rotating with $\omega_{d}/2$ can be expressed as (setting $\hbar=1$)
\begin{align}
\hat{H}_{\circlearrowleft}=&\hat{H}_{\textrm{c}}+\hat{H}_{\textrm{m}}+\hat{H}_{\textrm{om}}+\hat{H}_{\textrm{dr},\circlearrowleft}, \notag \\
\hat{H}_{\textrm{c}}=&\Delta_{c}\hat{a}_{\circlearrowleft}^{\dagger}\hat{a}_{\circlearrowleft}
+\Xi_{d}(e^{-i\theta_{d}}\hat{a}_{\circlearrowleft}^{\dagger2}+e^{i\theta_{d}}\hat{a}_{\circlearrowleft}^{2}),
\notag
\end{align}
\begin{align}
\hat{H}_{\textrm{m}}=&\dfrac{\omega_{m,1}}{2}(\hat{p}_{1}^{2}+\hat{q}_{1}^{2})
+\dfrac{\omega_{m,2}}{2}(\hat{p}_{2}^{2}+\hat{q}_{2}^{2})
+\lambda\hat{q}_{1}\hat{q}_{2},
\notag \\
\hat{H}_{\textrm{om}}=&-g_{1}\hat{a}_{\circlearrowleft}^{\dagger}\hat{a}_{\circlearrowleft}\hat{q}_{1}-g_{2}\hat{a}_{\circlearrowleft}^{\dagger}\hat{a}_{\circlearrowleft}\hat{q}_{2},
\notag \\
\hat{H}_{\textrm{dr},\circlearrowleft}=&i\varepsilon_{d,\circlearrowleft}(\hat{a}_{\circlearrowleft}^{\dagger}-\hat{a}_{\circlearrowleft}),
\label{eq_Ham1}
\end{align}
where $\hat{a}_{\circlearrowleft}$ ($\hat{a}_{\circlearrowleft}^{\dagger}$) is the annihilation (creation) operator of the CCW mode, while $\hat{q}_{j}$ and $\hat{p}_{j}$ are the dimensionless displacement and momentum operators of the $j$th mechanical mode, respectively. $\Delta_{c}=\omega_{c}-\omega_{d}/2$ is the optical detuning between the cavity mode and the coherent driving field. In Eq.\,(\ref{eq_Ham1}), $\hat{H}_{\textrm{c}}$ is the Hamiltonian of the cavity mode, in which the first term represents the free optical Hamiltonian and the second term describes the squeezing interaction with frequency $\omega_{d}$, amplitude $\Xi_{d}$, and phase $\theta_{d}$. In practice, one standard approach to generate such squeezing interaction is to excite the optical parametric down-conversion process in $\chi^{(2)}$ nonlinear materials via a two-photon pumping field~\cite{Safavi2013Nature}. As shown in Figs.\,\ref{Fig1}(b) and \ref{Fig1}(c), this intracavity squeezing interaction not only enhances the effective COM interaction, but also increases optical input noises simultaneously, thus leading to competing effects in quantum engineering~\cite{Jiao2024LPR}. To eliminate the negative influence, one can inject a broadband squeezed optical field along with the two-photon pumping field, which enables an effective suppression of optical input noises under phase-matched condition. Experimentally, this driving process can be implemented by mixing the two optical fields in a beam splitter (BS) [see Fig.\,\ref{Fig1}(a)], which will be further elaborated upon in the following discussion. $H_{\textrm{m}}$ is the Hamiltonian of the two mechanical modes, in which the first two terms are the free mechanical Hamiltonian and the third term denotes a phonon-hopping interaction with strength $\lambda$ between the two mechanical modes. It should be noted that, when the two mechanical eigenmodes have widely different resonant frequencies, the optomechanically hybridization effect becomes extremely weak~\cite{Shkarin2014PRL}, so that the phonon-hopping coupling could be negligible. To ensure the validity of the phonon-hopping coupling, we assume that the two mechanical modes in our system remain degenerate throughout the simulation. $H_{\textrm{om}}$ describes the nonlinear COM interactions between the cavity mode and the two mechanical modes, which are mediated by the optical radiation pressure exerted on the nanomechanical oscillators. In near-field COM systems, the COM coupling strength $g_{j}$ depends on the position $x_{0}$ where the mechanical oscillator is placed within the evanescent field, that is, $g_{j}(x_{0})=d\omega_{c}/dx|_{x=x_{0}}$. Owing to this spatial dependence of $g_{j}$, the COM coupling strength of each mechanical oscillator can be individually tailored by positioning the string at specific locations along the field gradient. $H_{\textrm{dr}}$ describes the Hamiltonian of the coherent driving field, in which $|\varepsilon_{d,\circlearrowleft}|=\sqrt{2\kappa P_{d}/\hbar\omega_{d}}$ denotes the field amplitude, with $P_{d}$ the input laser power and $\kappa$ the optical decay rate. It is worth noting that the total Hamiltonian is identical for the CW and CCW input cases, as light would experience the same optical and optomechanical nonlinearities in these counterpropagating modes. Therefore, for the CW input case, the Hamiltonian of the whole system can be directly obtained by altering the index of the corresponding parameters and operators from $\circlearrowleft$ to $\circlearrowright$ in Eq.\,(\ref{eq_Ham1}).

In our proposed scheme, the primary distinction between the CW and CCW input cases lies in the squeezing phase and strength of their injected squeezed vacuum reservoirs, which could induce asymmetric quantum noise flows in these counterpropagating modes. In the following, we discuss how to independently inject squeezed vacuum reservoirs to CW and CCW modes, and then derive the effective master equation for CCW input case as a specific example. As shown in Fig.\,\ref{Fig1}(a), we consider applying two different broadband squeezed optical fields to these counterpropagating modes, which are centred around frequency $\omega_{c}$ and characterized by squeezing strengths $r_{e,i}$ and reference phase angles $\theta_{e,i}$ ($i=\circlearrowleft,\circlearrowleft$), respectively. In practice, these broadband squeezed optical fields can act as two independent squeezed-vacuum reservoirs to CW and CCW modes~\cite{Kashiwazaki2020APL}. This is because the bandwidth of squeezed optical field can be up to gigahertz ($\textrm{GHz}$)~\cite{Murch2013Nature}, while the typical linewidth of WGM resonators is of the order of megahertz ($\textrm{MHz}$). In this case, the broadband squeezed optical field can be well approximated as having infinite bandwidth to WGM optical modes and be safely treated as a squeezed vacuum reservoir. Besides, the two mechanical modes are assumed to be coupled with two independent thermal reservoirs at the same bath temperature $T$. Then, in terms of the CCW input case, including the dissipations of the optical and mechanical modes, the dynamics of the total system is governed by the following Born-Markovian master equation~\cite{Zoller2000book}
\begin{align}
\dot{\rho}_{\circlearrowleft}=&i\big[\rho_{\circlearrowleft},\hat{H}_{\circlearrowleft}\big]+\frac{\kappa}{2}(N_{e,\circlearrowleft}+1)\mathcal{D}[\hat{a}_{\circlearrowleft}]\rho_{\circlearrowleft}+\frac{\kappa}{2}N_{e,\circlearrowleft}\mathcal{D}[\hat{a}_{\circlearrowleft}^{\dagger}]\rho_{\circlearrowleft}
\notag \\
&-\frac{\kappa}{2}M_{e,\circlearrowleft}\mathcal{G}[\hat{a}_{\circlearrowleft}]\rho_{\circlearrowleft}-\frac{\kappa}{2}M_{e,\circlearrowleft}^{\ast}\mathcal{G}[\hat{a}_{\circlearrowleft}^{\dagger}]\rho_{\circlearrowleft}
\notag \\
&-\sum_{j=1,2}\left(i\dfrac{\gamma_{m,j}}{2}[\hat{q}_{j},\{\hat{p}_{j},\rho_{\circlearrowleft}\}]+\gamma_{m,j}\bar{n}_{m,j}[\hat{q}_{j},[\hat{q}_{j},\rho_{\circlearrowleft}]]\right), \end{align}
where
\begin{align}
\mathcal{D}[\hat{a}_{\circlearrowleft}]\rho_{\circlearrowleft}&=2\hat{a}_{\circlearrowleft}\rho_{\circlearrowleft}\hat{a}_{\circlearrowleft}^{\dagger}-(\hat{a}_{\circlearrowleft}^{\dagger}\hat{a}_{\circlearrowleft}\rho_{\circlearrowleft}+\rho_{\circlearrowleft}\hat{a}_{\circlearrowleft}^{\dagger}\hat{a}_{\circlearrowleft}),
\notag \\
\mathcal{G}[\hat{a}_{\circlearrowleft}]\rho_{\circlearrowleft}&=2\hat{a}_{\circlearrowleft}\rho_{\circlearrowleft}\hat{a}_{\circlearrowleft}-(\hat{a}_{\circlearrowleft}\hat{a}_{\circlearrowleft}\rho_{\circlearrowleft}+\rho_{\circlearrowleft}\hat{a}_{\circlearrowleft}\hat{a}_{\circlearrowleft})
\end{align}
are the Lindblad operators, while $[\cdot,\cdot]$ and $\{\cdot,\cdot\}$ denote the commutator and anti-commutator, respectively. $\rho_{\circlearrowleft}$ is the density operator in the original picture. $\gamma_{m,j}$ is the mechanical damping rate of the $j$th mechanical mode, and $\bar{n}_{m,j}=1/[\exp(\omega_{m,j}/k_{B}T)-1]$ is the associated mean thermal phonon excitation number, with $k_{B}$ the Boltzmann constant. $N_{e,\circlearrowleft}=\sinh^{2}(r_{e,\circlearrowleft})$ and $M_{e,\circlearrowleft}=\cosh(r_{e,\circlearrowleft})\sinh(r_{e,\circlearrowleft})e^{i\theta_{e,\circlearrowleft}}$ describe the dissipation and the two-photon correlation of the cavity field caused by the squeezed-vacuum reservoir, respectively.

After performing the Bogoliubov transformation with a unitary operator, $U_{\circlearrowleft}(\eta_{d})=\exp[(-\eta_{d}\hat{a}_{\circlearrowleft}^{\dagger2}+\eta_{d}^{\ast}\hat{a}_{\circlearrowleft}^{2})/2]$, a squeezed optical mode $\hat{a}_{s,\circlearrowleft}$ can be introduced, i.e.,
\begin{align}
U_{\circlearrowleft}^{\dagger}(\eta_{d})\hat{a}_{\circlearrowleft}U_{\circlearrowleft}(\eta_{d})=\cosh(r_{d})\hat{a}_{s,\circlearrowleft}-e^{-i\theta_{d}}\sinh(r_{d})\hat{a}_{s,\circlearrowleft}^{\dagger},
\label{eq_bogo}
\end{align}
where $\eta_{d}=r_{d}e^{-i\theta_{d}}$ is the complex squeezing parameter, with a squeezing strength $r_{d}=(1/4)\ln[(\Delta_{c}+2\Xi_{d})/(\Delta_{c}-2\Xi_{d})]$ and a squeezing reference angle $\theta_{d}$. Hence, by dropping the constant terms, the effective Hamiltonian with respect to the CCW input case in the squeezing picture is derived as [see Appendix \ref{AppA} for more details]
\begin{align}
\hat{\tilde{H}}_{\circlearrowleft}
=&\omega_{s}\hat{a}_{s,\circlearrowleft}^{\dagger}\hat{a}_{s,\circlearrowleft}+\sum_{j=1,2}\left[\dfrac{\omega_{m,j}}{2}(\hat{p}_{j}^{2}+\hat{q}_{j}^{2})
-\zeta_{s,j}\hat{a}_{s,\circlearrowleft}^{\dagger}\hat{a}_{s,\circlearrowleft}\hat{q}_{j}\right.\notag \\
&\left.
+\dfrac{\zeta_{p,j}}{2}(e^{-i\theta_{d}}\hat{a}_{s,\circlearrowleft}^{\dagger2}+e^{i\theta_{d}}\hat{a}_{s,\circlearrowleft}^{2})\hat{q}_{j}-F_{j}\hat{q}_{j} \right]+\lambda\hat{q}_{1}\hat{q}_{2}\notag \\
&+i\varepsilon_{d,\circlearrowleft}\sinh(r_{d})(e^{-i\theta_{d}}\hat{a}_{s,\circlearrowleft}^{\dagger}-e^{i\theta_{d}}\hat{a}_{s,\circlearrowleft})\notag \\
&+i\varepsilon_{d,\circlearrowleft}\cosh(r_{d})(\hat{a}_{s,\circlearrowleft}^{\dagger}-\hat{a}_{s,\circlearrowleft}),
\label{eq_Ham2}
\end{align}
where $\omega_{s}=(\Delta_{c}-2\Xi_{d})\exp(2r_{d})$ is the effective resonance frequency of the squeezed optical mode. $\zeta_{s,j}=g_{j}\cosh(2r_{d})$ and $\zeta_{p,j}=g_{j}\sinh(2r_{d})$ are the effective COM coupling strength for the $j$th mechanical mode induced by the radiation pressure and the parametric amplification, respectively. $F_{j}=g_{j}\sinh^{2}(r_{d})$ denotes the strength of the constant mechanical driving force exerted on the $j$th mechanical mode, which is caused by the parametric amplification. Correspondingly, the effective master equation in the squeezing picture is derived as [see Appendix \ref{AppB} for more details]
\begin{align}
\dot{\tilde{\rho}}_{\circlearrowleft}=&U^{\dagger}(\eta_{d})\dot{\rho}_{\circlearrowleft}U(\eta_{d})
\notag \\
=&i\big[\tilde{\rho}_{\circlearrowleft},\hat{\tilde{H}}_{\circlearrowleft}\big]\!+\!\frac{\kappa}{2}(N_{s,\circlearrowleft}+1)\mathcal{D}[\hat{a}_{s,\circlearrowleft}]\tilde{\rho}_{\circlearrowleft}
\!+\!\frac{\kappa}{2}N_{s,\circlearrowleft}\mathcal{D}[\hat{a}_{s,\circlearrowleft}^{\dagger}]\tilde{\rho}_{\circlearrowleft}
\notag \\
&-\frac{\kappa}{2}M_{s,\circlearrowleft}\mathcal{G}[\hat{a}_{s,\circlearrowleft}]\tilde{\rho}_{\circlearrowleft}
-\frac{\kappa}{2}M_{s,\circlearrowleft}^{\ast}\mathcal{G}[\hat{a}_{s,\circlearrowleft}]\tilde{\rho}_{\circlearrowleft}
\notag \\
&-\sum_{j=1,2}\left(i\dfrac{\gamma_{m,j}}{2}[\hat{q}_{j},\{\hat{p}_{j},\tilde{\rho}_{\circlearrowleft}\}]+\gamma_{m,j}\bar{n}_{m,j}[\hat{q}_{j},[\hat{q}_{j},\tilde{\rho}_{\circlearrowleft}]]\right),
\label{eq_effmeq}
\end{align}
where
\begin{align}
\nonumber
N_{s,\circlearrowleft}=&\sinh^{2}(r_{d})\cosh^{2}(r_{e,\circlearrowleft})+\cosh^{2}(r_{d})\sinh^{2}(r_{e,\circlearrowleft})
\notag \\
&+\dfrac{1}{2}\cos\delta\theta\sinh(2r_{d})\sinh(2r_{e,\circlearrowleft}),
\notag\\
M_{s,\circlearrowleft}\!=&e^{i\theta_{d}}\![\sinh(r_{d})\!\cosh(r_{e,\circlearrowleft})\!+\!e^{i\delta\theta_{\circlearrowleft}}\!\cosh(r_{d})\!\sinh(r_{e,\circlearrowleft})\!]
\notag\\
&\times\![\cosh(r_{d})\!\cosh(r_{e,\circlearrowleft})\!+\!e^{-i\delta\theta_{\circlearrowleft}}\!\sinh(r_{d})\!\sinh(r_{e,\circlearrowleft})],
\label{eq_NsMs}
\end{align}
where $N_{s,\circlearrowleft}$ and $M_{s,\circlearrowleft}$ denote the effective thermal noise and the two-photon-correlation strength of the CCW mode, respectively. $\tilde{\rho}_{\circlearrowleft}=U^{\dagger}(\eta_{d})\rho_{\circlearrowleft}U(\eta_{d})$ is the density operator in the squeezing picture. $\delta\theta_{\circlearrowleft}\equiv\theta_{e,\circlearrowleft}-\theta_{d}$ ($\delta r_{\circlearrowleft}\equiv r_{e,\circlearrowleft}-r_{d}$) is the squeezing phase (strength) difference between the CCW mode and the squeezed-vacuum reservoir. Note that both CW and CCW modes are coupled with squeezed vacuum reservoirs and their only distinction lies in the squeezing phase and strength of these reservoirs, the effective master equation for the CW input case can thus be directly obtained from Eq.\,(\ref{eq_effmeq}) by altering the index of the corresponding parameters and operators from $\circlearrowleft$ to $\circlearrowright$. From Eq.\,(\ref{eq_NsMs}), it is obvious that in the absence of the squeezed-vacuum reservoir (i.e., $r_{e,i}=0$ and $\theta_{e,i}=0$), $N_{s,i}=\sinh^{2}(r_{d})$ and $M_{s,i}=\cosh(r_{d})\sinh(r_{d})e^{i\theta_{d}}$ would be inevitably amplified when the squeezing strength $r_{d}$ increases, leading to an intracavity-squeezing-enhanced light-reservoir interaction. On the other hand, in the presence of the squeezed-vacuum reservoir (i.e., $r_{e,i}\neq0$ and $\theta_{e,i}\neq0$), it is seen that these additional thermalization noises become controllable through $\delta r_{i}$ and $\delta\theta_{i}$ [see Fig.\,\ref{Fig1}(c)], which can be coherently tuned by the squeezing parameters $r_{e,i}$ and $\theta_{e,i}$. In particular, when choosing $\delta r_{i}=0$ and $\delta\theta_{i}=\pm n\pi (n=1,3,5...)$, both $N_{s,i}$ and $M_{s,i}$ can be completely eliminated, i.e., $N_{s,i}=M_{s,i}=0$, yielding the phase-matched condition. In contrast, for other values of $\delta r_{i}$ and $\delta\theta_{i}$, $N_{s,i}$ and $M_{s,i}$ have nonzero values, corresponding to the phase-mismatched condition. Then, by selectively choosing different squeezing parameters $r_{e,i}$ and $\theta_{e,i}$ for the CW and CCW modes, $N_{s,i}$ and $M_{s,i}$ can thus be controlled in an asymmetric way, which leads to asymmetric noise flows in these counterpropagating modes. In the following, as a specific example, we consider the case where the squeezed optical mode $\hat{a}_{s,\circlearrowleft}$ is phase-matched with its squeezed vacuum reservoir, resulting in $N_{s,\circlearrowleft}=M_{s,\circlearrowleft}=0$, whereas the condition is phase-mismatched for $\hat{a}_{s,\circlearrowright}$, leading to $N_{s,\circlearrowright}\neq0$ and $M_{s,\circlearrowright}\neq0$. It is worth emphasizing that this quantum-noise-induced asymmetry can lead to differences in the quantum statistical properties of the CW and CCW modes [see Fig.\,\ref{Fig1}(d)], which forms the basis for the implementation of the phase-selective entanglement and asymmetric EPR steering in this configuration. The physics behind the result in Fig.\,\ref{Fig1}(d) will be discussed in detail in Sec.\,\ref{secIII}.

To quantitatively investigate the behavior of quantum entanglement and EPR steering, one should solve the system dynamics by employing the effective mater equation (\ref{eq_effmeq}). However, due to the nonlinear COM interactions in Hamiltonian (\ref{eq_Ham2}), the system dynamics is difficult to be directly solved. In order to deal with this problem, one can linearize the Hamiltonian by expanding each operator as a sum of its steady-state mean value and a small quantum fluctuation around it under the condition of strong coherent optical driving, i.e., $\hat{a}_{s,i}=\bar{a}_{s,i}+\delta\hat{a}_{s,i}$, $\hat{q}_{j}=\bar{q}_{j}+\delta\hat{q}_{j}$, $\hat{p}_{j}=\bar{p}_{j}+\delta\hat{p}_{j}$. Since the linearization procedure is identical for the CW and CCW input cases, we here only present the linearization of the effective mater equation (\ref{eq_effmeq}) regarding the CCW input case. By using the master equation (\ref{eq_effmeq}), the steady-state mean values of the optical and mechanical modes with respect to the CCW input case are derived as
\begin{align}
\bar{a}_{s,\circlearrowleft}=&-\dfrac{(i\Delta_{s}-\dfrac{\kappa}{2})A_{1}+iA_{2}\beta_{p}}{(\Delta_{s}^2+\dfrac{\kappa^{2}}{4})-\beta_{p}^{2}}\varepsilon_{d,\circlearrowleft},
\notag \\
\bar{q}_{1}=&\dfrac{\omega_{m,2}B_{1,\circlearrowleft}-\lambda B_{2,\circlearrowleft}}{\omega_{m,1}\omega_{m,2}-\lambda^{2}},~~\bar{p}_{1}=0,
\notag \\
\bar{q}_{2}=&\dfrac{\omega_{m,1}B_{2,\circlearrowleft}-\lambda B_{1,\circlearrowleft}}{\omega_{m,1}\omega_{m,2}-\lambda^{2}},~~\bar{p}_{2}=0,
\label{Eq_ss}
\end{align}
with
\begin{align}
\Delta_{s} &= \omega_{s}-\beta_{s}, \notag \\
\alpha_{s,\circlearrowleft} &= e^{-i\theta_{d}}\bar{a}_{s,\circlearrowleft}^{\ast2}+e^{i\theta_{d}}\bar{a}_{s,\circlearrowleft}^{2}, \notag\\
\beta_{s} &= \zeta_{s,1}\bar{q}_{1}+\zeta_{s,2}\bar{q}_{2}, \notag \\
\beta_{p} &= \zeta_{p,1}\bar{q}_{1}+\zeta_{p,2}\bar{q}_{2}, \notag
\end{align}
\begin{align}
A_{1} &= \cosh(r_{d})+\sinh(r_{d})e^{-i\theta_{d}}, \notag \\
A_{2} &= \cosh(r_{d})e^{-i\theta_{d}}+\sinh(r_{d}), \notag \\
B_{1,\circlearrowleft} &= \zeta_{s,1}|\bar{a}_{s,\circlearrowleft}|^{2}-\dfrac{\zeta_{p,1}}{2}\alpha_{s,\circlearrowleft}+F_{1}, \notag \\
B_{2,\circlearrowleft} &= \zeta_{s,2}|\bar{a}_{s,\circlearrowleft}|^{2}-\dfrac{\zeta_{p,2}}{2}\alpha_{s,\circlearrowleft}+F_{2}.
\end{align}
Equation\,(\ref{Eq_ss}) indicates that the field amplitude $\bar{a}_{s,\circlearrowleft}$ is not only dependent on the strength of the coherent driving field, but also relies on the squeezing strength and reference angle of the squeezed optical mode. This allows us to regulate the COM coupling strength in a controllable way by adjusting the squeezing parameters. Moreover, it is also seen that the mean values of the mechanical displacement $\bar{q}_{j}$ are coupled to each other by a factor of $\lambda$, which is due to the phonon hopping process between the two mechanical modes.

Then, by substituting the expansions of the quantum operators into the Hamiltonian (\ref{eq_Ham2}), one can directly derive the linearized Hamiltonian as
\begin{align}
\hat{\tilde{H}}_{lin,\circlearrowleft}
=&\Delta_{s}\hat{a}_{s,\circlearrowleft}^{\dagger}\hat{a}_{s,\circlearrowleft}+\sum_{j=1,2}\left[\dfrac{\omega_{m,j}}{2}(\hat{p}_{j}^{2}+\hat{q}_{j}^{2})\right.\notag\\
&\left.-(\Lambda_{j}\hat{a}_{s,\circlearrowleft}^{\dagger}+\Lambda_{j}^{\ast}\hat{a}_{s,\circlearrowleft})\hat{q}_{j}\right]+\lambda\hat{q}_{1}\hat{q}_{2},
\label{eq_Hlin}
\end{align}
and the associated master equation as
\begin{align}
\dot{\tilde{\rho}}_{\circlearrowleft}
\!=&i\!\left[\tilde{\rho}_{\circlearrowleft},\tilde{H}_{lin,\circlearrowleft}\right]
\notag \\
&+\!\frac{\kappa}{2}(N_{s,\circlearrowleft}+1)\mathcal{D}[\hat{a}_{s,\circlearrowleft}]\tilde{\rho}_{\circlearrowleft}
\!+\!\frac{\kappa}{2}N_{s,\circlearrowleft}\mathcal{D}[\hat{a}_{s,\circlearrowleft}^{\dagger}]\tilde{\rho}_{\circlearrowleft}
\notag \\
&-\frac{\kappa}{2}M_{s,\circlearrowleft}\mathcal{G}[\hat{a}_{s,\circlearrowleft}]\tilde{\rho}_{\circlearrowleft}
-\frac{\kappa}{2}M_{s,\circlearrowleft}^{\ast}\mathcal{G}[\hat{a}_{s,\circlearrowleft}]\tilde{\rho}_{\circlearrowleft}
\notag \\
&-\sum_{j=1,2}\left(i\dfrac{\gamma_{m,j}}{2}[\hat{q}_{j},\{\hat{p}_{j},\tilde{\rho}_{\circlearrowleft}\}]+\gamma_{m,j}\bar{n}_{m,j}[\hat{q}_{j},[\hat{q}_{j},\tilde{\rho}_{\circlearrowleft}]]\right),
\label{eq_meq3}
\end{align}
where
\begin{align}
\Lambda_{j}=G_{j}\cosh(2r_{d})-G_{j}^{\ast}\sinh(2r_{d})e^{-i\theta_{d}} \label{eq_coupling}
\end{align}
is the effective COM coupling rate, with $G_{j}=g_{j}\bar{a}_{s,\circlearrowleft}$. By defining an enhancement factor $\Pi_{j}=|\Lambda_{j}/G_{j}|$, it is seen that with proper squeezing parameters, the COM coupling rate can be effectively enhanced [see Fig.\,\ref{Fig1}(b)]. Note that for notational convenience, we have neglected the symbol $``\delta"$ in the expression of quantum fluctuation operators in Eqs.\,(\ref{eq_Hlin}) and (\ref{eq_meq3}). We also emphasize that in the weak COM coupling regime, the steady-state mean value of the optical mode is much larger than those of the mechanical modes, i.e., $|\bar{a}_{s,\circlearrowleft}|\gg|\bar{q}_{j}|$. Under this condition, we have ignored the terms $\hat{a}_{s,\circlearrowleft}^{2}$ and $\hat{a}_{s,\circlearrowleft}^{\dagger2}$ in Hamiltonian\,(\ref{eq_Hlin}), whose coefficient $\beta_{p}$ is dominated by $\bar{q}_{j}$. In the following discussions, for ensuring the validity of Hamiltonian\,(\ref{eq_Hlin}), the system parameters have been strictly restricted to satisfy the condition of weak COM coupling.

\begin{figure*}[htbp]
\centering
\includegraphics[width=0.96\textwidth]{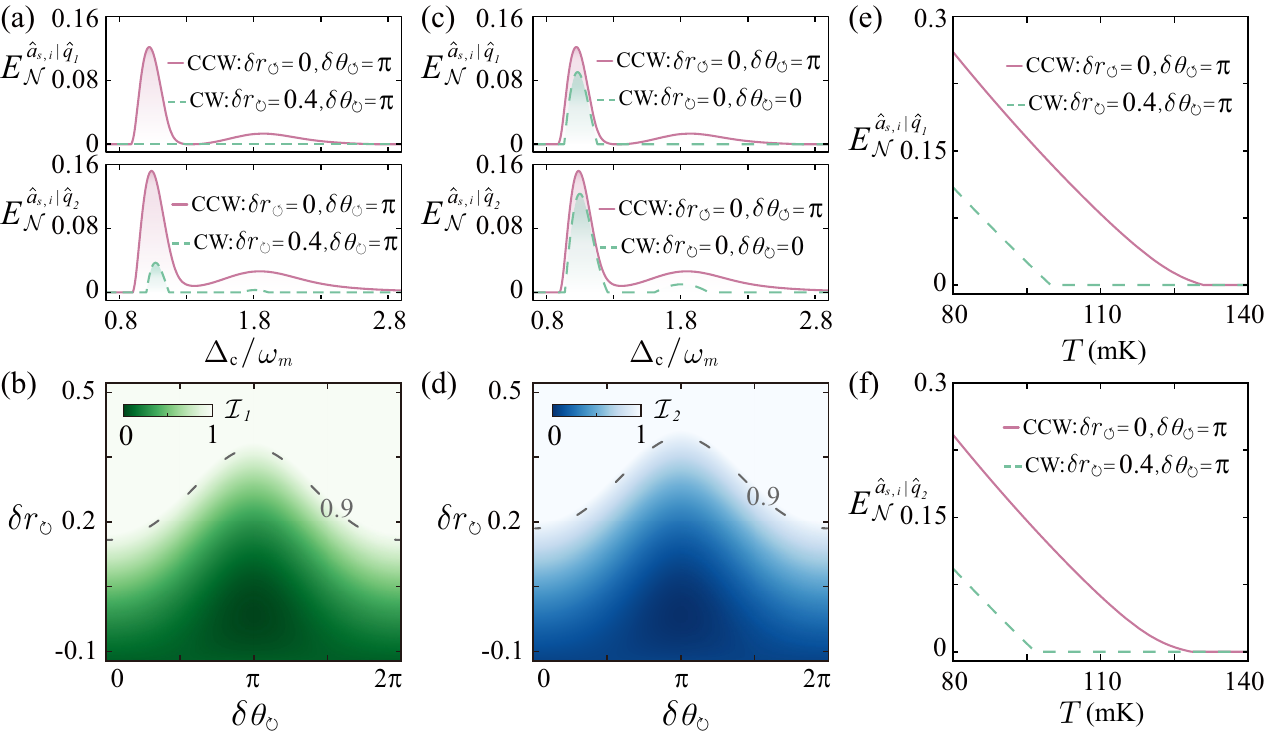}
\caption{\label{Fig2}Phase-selective generation of optomechanical entanglement. (a),(c) Plot of the logarithmic negativity $E_{\mathcal{N}}^{\hat{\textit{a}}_{\textit{s},\textit{i}}|\hat{\textit{q}}_{\textit{1}}}$ and $E_{\mathcal{N}}^{\hat{\textit{a}}_{\textit{s},\textit{i}}|\hat{\textit{q}}_{\textit{2}}}$ versus the scaled optical detuning $\Delta_{c}/\omega_{m}$ for opposite optical input directions with $r_{d}=0.1$ and $\theta_{d}=\pi$. (b),(d) Density plot of $\mathcal{I}_{1}$ and $\mathcal{I}_{2}$ as functions of $\delta r_{\circlearrowright}$ and $\delta\theta_{\circlearrowright}$, with $\delta r_{\circlearrowleft}=0$ and $\delta\theta_{\circlearrowleft}=\pi$. (e),(f) Plot of the logarithmic negativity $E_{\mathcal{N}}^{\hat{\textit{a}}_{\textit{s},\textit{i}}|\hat{\textit{q}}_{\textit{1}}}$ and $E_{\mathcal{N}}^{\hat{\textit{a}}_{\textit{s},\textit{i}}|\hat{\textit{q}}_{\textit{2}}}$  versus the mechanical environment temperature $T$ for opposite optical input directions with $r_{d}=0.1$, $\theta_{d}=\pi$, and $\Delta_{c}=1$.}
\end{figure*}

Since the system dynamics is linearized now and the input noises for the optical and mechanical modes are Gaussian, the steady state of the system, independently of any initial state, could eventually evolve into a tripartite zero-mean Gaussian state, whose statistic is fully characterized by a $6\times6$ covariance matrix (CM) $V$ with its matrix element
\begin{align}
V_{kl}=\langle \psi_{k}\psi_{l}+\psi_{l}\psi_{k}\rangle/2,~(k,l=1,2,\ldots,6).
\label{eq_Vkl}
\end{align}
Here $\psi=(\hat{X}_{s,\circlearrowleft},\hat{Y}_{s,\circlearrowleft},\hat{q}_{1},\hat{p}_{1},\hat{q}_{2},\hat{p}_{2})^{T}$ is the vector of optical and mechanical quadrature operators, with its components defined by
\begin{align}
&\hat{X}_{s,\circlearrowleft}=\dfrac{1}{\sqrt{2}}\left(\hat{a}_{s,\circlearrowleft}^{\dagger}+\hat{a}_{s,\circlearrowleft}\right),
~\hat{Y}_{s,\circlearrowleft}=\dfrac{i}{\sqrt{2}}\left(\hat{a}_{s,\circlearrowleft}^{\dagger}-\hat{a}_{s,\circlearrowleft}\right).
\label{eq_quad}
\end{align}
Employing the linearized master equation\,(\ref{eq_meq3}), one can obtain the dynamics of arbitrary quantum correlation between the optical and mechanical quadrature operators in the CM $V$. In terms of optical bosonic operator, we define the following quantum correlation functions
\begin{align}
&x_{1}=\langle\hat{a}_{s,\circlearrowleft}^{\dagger}\hat{a}_{s,\circlearrowleft}\rangle,
~~x_{2}=\langle\hat{a}_{s,\circlearrowleft}\hat{a}_{s,\circlearrowleft}^{\dagger}\rangle,
~~x_{3}=\langle\hat{q}_{1}\hat{q}_{1}\rangle,
\notag \\
&x_{4}=\langle\hat{p}_{1}\hat{p}_{1}\rangle,
~~x_{5}=\langle\hat{q}_{2}\hat{q}_{2}\rangle,
~~x_{6}=\langle\hat{p}_{2}\hat{p}_{2}\rangle,
\notag \\
&x_{7}=\langle\hat{a}_{s,\circlearrowleft}\hat{a}_{s,\circlearrowleft}\rangle,
~~x_{8}=\langle\hat{a}_{s,\circlearrowleft}^{\dagger}\hat{a}_{s,\circlearrowleft}^{\dagger}\rangle,
~~x_{9}=\langle\hat{q}_{1}\hat{p}_{1}\rangle,
\notag \\
&x_{10}=\langle\hat{p}_{1}\hat{q}_{1}\rangle,
~~x_{11}=\langle\hat{q}_{2}\hat{p}_{2}\rangle,
~~x_{12}=\langle\hat{p}_{2}\hat{q}_{2}\rangle,
\notag \\
&x_{13}=\langle\hat{a}_{s,\circlearrowleft}\hat{q}_{1}\rangle,
~~x_{14}=\langle\hat{a}_{s,\circlearrowleft}^{\dagger}\hat{q}_{1}\rangle,
~~x_{15}=\langle\hat{a}_{s,\circlearrowleft}\hat{p}_{1}\rangle,
\notag \\
&x_{16}=\langle\hat{a}_{s,\circlearrowleft}^{\dagger}\hat{p}_{1}\rangle,
~~x_{17}=\langle\hat{a}_{s,\circlearrowleft}\hat{q}_{2}\rangle,
~~x_{18}=\langle\hat{a}_{s,\circlearrowleft}^{\dagger}\hat{q}_{2}\rangle,
\notag \\
&x_{19}=\langle\hat{a}_{s,\circlearrowleft}\hat{p}_{2}\rangle,
~~x_{20}=\langle\hat{a}_{s,\circlearrowleft}^{\dagger}\hat{p}_{2}\rangle,
~~x_{21}=\langle\hat{q}_{1}\hat{q}_{2}\rangle,
\notag \\
&x_{22}=\langle\hat{p}_{1}\hat{p}_{2}\rangle,
~~x_{23}=\langle\hat{q}_{1}\hat{p}_{2}\rangle,
~~x_{24}=\langle\hat{q}_{2}\hat{p}_{1}\rangle.
\end{align}
By grouping together these quantum correlations in a vector $X=(x_{1},x_{2},\ldots,x_{24})^{T}$ and employing the linearized effective master equation (\ref{eq_meq3}), one can obtain its time evolution equation as
\begin{align}
\dfrac{d}{dt}X=A\cdot X+b, \label{eq_eom}
\end{align}
where
\begin{align}
b=&\left[\kappa N_{s},\kappa(N_{s}+1),0,2\gamma_{m,1}\bar{n}_{m,1},0,2\gamma_{m,2}\bar{n}_{m,2},\kappa M_{s}^{\ast},\right.
\notag \\
&\left.\kappa M_{s},0,0,0,0,0,0,0,0,0,0,0,0,0,0,0,0\right]^{T},
\end{align}
is a vector involving the correlation functions for the input quantum noises. Here
the exact expression for the coefficient matrix $A$ and the detailed derivation process of the time evolution equation\,(\ref{eq_eom}) are too cumbersome, and, for convenience, we have reported them in Appendix \ref{AppC}. Notably, by numerically solving Eq.\,(\ref{eq_eom}) and using the relations between bosonic operators and quadrature operators, one can directly obtain the steady-state CM $V$.

Regarding the verification of bipartite and tripartite entanglement, we adopt the logarithmic negativity $E_{\mathcal{N}}$ and the minimum residual contangle $\mathcal{R}^{\textrm{min}}_{\tau}$ as the quantitative entanglement measures, respectively, which are defined based on specifying the positivity of the partial transpose of the CM $V$. For continuous-variable (CV) bipartite Gaussian state, the logarithmic negativity $E_{\mathcal{N}}^{\mu|\nu}$ is defined as~\cite{Adesso2004pra}
\begin{align}
E_{\mathcal{N}}^{\mu|\nu}=&\max\,\big[0,-\ln(2\eta_{0}^{-})\big],
\label{En}
\end{align}
where $\eta_{0}^{-}\!\equiv\!2^{-1/2}\{\Sigma(V_{\mu\nu})-[\Sigma(V_{\mu\nu})^{2}-4\det\!V_{\mu\nu}]^{1/2}\}^{1/2}$, with $\Sigma(V_{\mu\nu})\!\equiv\!\det\mathcal{A}_{\mu}+\det\mathcal{B}_{\nu}-2\det\mathcal{C}_{\mu\nu}$, is the minimum symplectic eigenvalue of the partial transpose of a reduced $4\times4$ CM $V_{\mu\nu}$, with $\mu$ and $\nu$ describing the selected modes under consideration ($\mu,\,\nu=\hat{a}_{s,\circlearrowright},\hat{a}_{s,\circlearrowleft},\hat{q}_{1},\hat{q}_{2}$). The reduced CM $V_{\mu\nu}$ contains the entries of $V$, and it can be obtained by selecting the rows and columns associated with modes $\mu$ and $\nu$ from $V$, whose $ 2\times2 $ block form is given by
\begin{align}
V_{\mu\nu}=\left(
\begin{matrix}
\mathcal{A}_{\mu}&\mathcal{C}_{\mu\nu}\\
\mathcal{C}_{\mu\nu}^{\textit{T}}&\mathcal{B}_{\nu}
\end{matrix}
\right).\label{reducedCM}
\end{align}
Equation\,(\ref{En}) quantifies how much the positivity of the partial transpose condition for separability is violated for the considered Gaussian states, which is equivalent to Simon's necessary and sufficient entanglement nonpositive partial transpose criterion (or the related Peres-Horodecki criterion)~\cite{Simon2000PRL}. Note that the selected modes $\mu$ and $\nu$ gets entangled if and only if $\eta_{0}^{-}<1/2$, where $E_{\mathcal{N}}$ has a nonzero value.
\begin{figure}[htbp]
\centering
\includegraphics[width=0.48\textwidth]{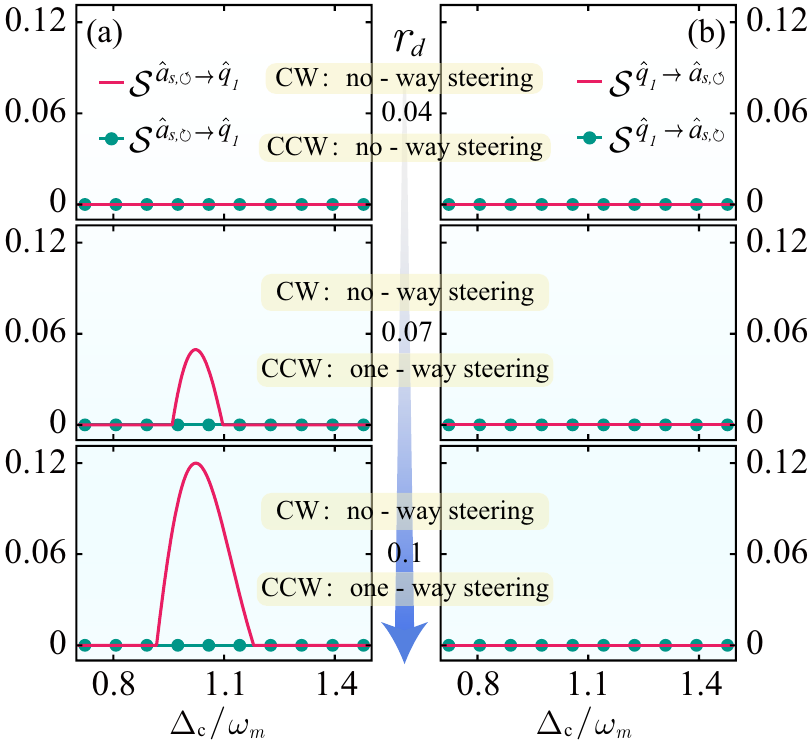}
\caption{\label{Fig3}Phase-selective generation of asymmetric optomechanical EPR steering. The quantum steerability $\mathcal{S}^{\hat{a}_{s,i}\rightarrow\hat{q}_{1}}$ (a) and $\mathcal{S}^{\hat{q}_{1}\rightarrow\hat{a}_{s,i}}$ (b) for opposite optical input directions are plotted as functions of the scaled optical detuning $\Delta_{c}/\omega_{m}$ under different squeezing parameter $r_{d}$, with $\delta r_{\circlearrowleft}=\textrm{0}$, $\delta r_{\circlearrowright}=\textrm{0.4}$, and $\delta\theta_{\circlearrowleft}=\delta\theta_{\circlearrowright}=\pi$. With the increase of $r_{d}$, EPR steering can be driven from no-way regime to one-way regime for the CCW input case but not for the CW input case, implying that the asymmetry of EPR steering are effectively controlled by the phase-matched condition.}
\end{figure}

The minimum residual contangle, $\mathcal{R}^{\textrm{min}}_{\tau}$, which provides a \textit{bona fide} quantification of CV tripartite entanglement, is defined as~\cite{Adesso2007JPA}
\begin{align}
\mathcal{R}^{\textrm{min}}_{\tau}=&\min_{(r,s,t)}\,\big[E^{r|st}_{\tau}-E^{r|s}_{\tau}-E^{r|t}_{\tau}\big],
\end{align}
where $(r,s,t)$ denotes all the possible permutations of the three-mode indexes. $E_{\tau}$ is the one-mode-versus-(one- or two-mode) contangle, which can be defined by a proper entanglement monotone, e.g., the squared logarithmic negativity. Based on Eq.\,(\ref{En}), the one-mode-versus-one-mode contangle $E^{r|s}_{\tau}$ and $E^{r|t}_{\tau}$ can be directly obtained by employing its definition, namely, $E_{\tau}^{r|s}\equiv[E_{\mathcal{N}}^{r|s}]^{2}$ and $E_{\tau}^{r|t}\equiv[E_{\mathcal{N}}^{r|t}]^{2}$. However, when calculating the one-mode-versus-two-modes contangle $E^{r|st}_{\tau}$, one must alter the basic definition of Eq.\,(\ref{En}) by rewriting the definition of $\eta_{0}^{-}$ as given by $\eta_{0}^{-}\equiv\min\,[\textrm{eig}|i\Omega_{3}\tilde{V}_{r|st}|]$, where $\eta_{0}^{-}$ becomes the minimum symplectic eigenvalue of the partial transpose of a $6\times6$ CM $V_{r|st}$, with $\Omega_{3}=\oplus_{k=1}^{3}i\sigma_{y}$ and $\sigma_{y}$ the y-Pauli matrix. $\tilde{V}_{r|st}$ corresponds to the partial transpose of $V_{r|st}$, which connects to $V_{r|st}$ with the relation of $\tilde{V}_{r|st}=P_{r|st}VP_{r|st}$, and $P_{r|st}=\textrm{diag}(1,-1,1,1,1,1)$ is the partial transposition matrix. In terms of $E^{s|rt}_{\tau}$ and $E^{t|sr}_{\tau}$, the corresponding partial transpose matrices are given by $P_{s|rt}=\textrm{diag}(1,1,1,-1,1,1)$ and $P_{t|sr}=\textrm{diag}(1,1,1,1,1,-1)$, respectively. In addition, according to the Coffman-Kundu-Wootters monogamy inequality for quantum entanglement, we also note that the residual contangle is required to satisfy the following monogamy condition, i.e., $E^{r|st}_{\tau}-E^{r|s}_{\tau}-E^{r|t}_{\tau}\geq0$, which means that the bipartite entanglement between the partition $r$ and the remaining two partitions $st$ is never smaller than the sum of the $r|s$ and $r|t$ bipartite entanglements in the reduced states. As such, if there are nonzero values of the minimum residual contangle, i.e., $\mathcal{R}_{\tau}^{\textrm{min}}>0$, one can verify that the full tripartite entanglement is present for the CV system.

Besides, EPR steering is an asymmetric form of quantum entanglement, where a bipartite Gaussian state characterized by CM $V_{\mu\nu}$ may be steerable from mode $\mu$ to mode $\nu$, but not vice versa. To study the quantum steerability from mode $\mu$ to mode $\nu$ or from mode $\nu$ to mode $\mu$, we adopt an intuitive and computable quantification of EPR steering~\cite{Kogias2015PRL}, which are defined as
\begin{align}
\mathcal{S}^{\mu\rightarrow\nu}=&\max\left[0,\dfrac{1}{2}\ln\dfrac{\det\mathcal{A}_{\mu}}{4\det V_{\mu\nu}}\right], \notag\\
\mathcal{S}^{\nu\rightarrow\mu}=&\max\left[0,\dfrac{1}{2}\ln\dfrac{\det\mathcal{B}_{\nu}}{4\det V_{\mu\nu}}\right].
\end{align}
The quantity $\mathcal{S}$ is a monotone under Gaussian local operations and classical communication (LOCC) and the larger value of $\mathcal{S}$ implies the stronger Gaussian steerability. The steerability between mode $\mu$ and mode $\nu$ can be classified into three cases. For $\mathcal{S}^{\mu\rightarrow\nu}=\mathcal{S}^{\nu\rightarrow\mu}=0$, the state is non-steerable in any direction, which corresponds to the case of no-way steering. For $\mathcal{S}^{\mu\rightarrow\nu}>0$ and $\mathcal{S}^{\nu\rightarrow\mu}>0$, the state is steerable in both directions $\mu\rightarrow\nu$ and $\nu\rightarrow\mu$, which corresponds to the case of two-way steering. Finally, for $\mathcal{S}^{\mu\rightarrow\nu}>0$ and $\mathcal{S}^{\nu\rightarrow\mu}=0$ or $\mathcal{S}^{\nu\rightarrow\mu}>0$ and $\mathcal{S}^{\mu\rightarrow\nu}=0$, the state is only steerable in one certain direction $\mu\rightarrow\nu$ or $\nu\rightarrow\mu$, which corresponds to the case of one-way steering and reflects the asymmetric nature of entangled state.

\section{Phase-selective manipulation of quantum entanglement and EPR steering}\label{secIII}
First, we investigate the phase-dependent behaviors of optomechanical entanglement and the associated asymmetric EPR steering in Figs.\,\ref{Fig2} and \ref{Fig3}. To evaluate the entanglement measures, one can numerically calculate the steady-state CM $V$ by solving the dynamical equation (\ref{eq_eom}) with the following parameters: $\kappa/2\pi=4.9\,\textrm{MHz}$, $\omega_{m}/2\pi=16\,\textrm{MHz}$, $Q_{m}\equiv\omega_{m}/\gamma_{m}=10^{5}$, $G_{1}/2\pi=0.16\,\textrm{MHz}$, $G_{2}/2\pi=0.21\,\textrm{MHz}$, $\lambda/2\pi=0.32\,\textrm{MHz}$, and $\bar{n}_{m}=\textrm{100}$. Here, for ensuring the validity of the phonon-hopping interaction, we have supposed that the fundamental parameters of the two mechanical modes are identical, i.e, $\omega_{m,1}=\omega_{m,2}=\omega_{m}$, $\gamma_{m,1}=\gamma_{m,2}=\gamma_{m}$, $\bar{n}_{m,1}=\bar{n}_{m,2}=\bar{n}_{m}$. We note that these selected parameters ensure the condition required by the resolved sideband regime and mechanical ground-state cooling, i.e., $\kappa/\omega_{m}<1$, which are feasible for current experimental techniques~\cite{Anetsberger2009NP} and convenient for producing optomechanical entanglement and asymmetric EPR steering. In terms of the mean thermal phonon excitation number $\bar{n}_{m}$, it corresponds to a bath temperature of $T\backsimeq73\,\textrm{mK}$, which is achievable with the use of a dilution refrigerator. Moreover, since the effective COM coupling strength can be exponentially enhanced by the intracavity squeezing [see Fig.\,\ref{Fig1}(b)], we have chosen a much smaller value for $G_{j}$ in our simulation, whose typical value can reach up to $G_{j}/2\pi\backsim3.8\,\textrm{MHz}$ with a laser input power of $P=65\,\mu\textrm{W}$ in the near-field cavity optomechanical system ~\cite{Anetsberger2009NP}. And the asymmetric COM coupling strength for the two mechanical modes can be exmperimentally operated by placing the nanomechanical oscillators at different positions along the field gradient.

In Fig.\,\ref{Fig2}, we present how to achieve phase-selective generation of quantum entanglement regarding the bipartition of optical and mechanical modes. Figures\,\ref{Fig2}(a) and \ref{Fig2}(c) show the logarithmic negativity $E_{\mathcal{N}}^{\hat{a}_{s,i}|\hat{q}_{1}}$ and $E_{\mathcal{N}}^{\hat{a}_{s,i}|\hat{q}_{2}}$ for opposite optical input directions as functions of the scaled optical detuning $\Delta_{c}/\omega_{m}$. Here $E_{\mathcal{N}}^{\hat{a}_{s,i}|\hat{q}_{j}}$ corresponds to the logarithmic negativity of the bipartition of the squeezed optical mode $i$ and the mechanical mode $j$. As discussed in previous section, we have assumed the fulfillment of the phase-matched condition for the CCW input case [i.e., $\delta r_{\circlearrowleft}=\textrm{0}$ and $\delta\theta_{\circlearrowleft}=\pm n\pi (n=1,3,5...)$], whereas such condition is phase-mismatched for the CW input case [i.e., $\delta r_{\circlearrowright}\neq\textrm{0}$ and $\delta\theta_{\circlearrowright}\neq\pm n\pi~(n=1,3,5...)$]. Under this circumstance, it is seen that the profiles of $E_{\mathcal{N}}^{\hat{a}_{s,\circlearrowleft}|\hat{q}_{1}}$ and $E_{\mathcal{N}}^{\hat{a}_{s,\circlearrowleft}|\hat{q}_{2}}$ for the CCW input case are characterized by a sharp peak around the optical detuning at COM resonance $\Delta_{c}/\omega_{m}\sim\textrm{1}$ (purple solid curve). In contrast, when reversing the coherent driving field, the values of the logarithmic negativity $E_{\mathcal{N}}^{\hat{a}_{s,\circlearrowright}|\hat{q}_{1}}$ and $E_{\mathcal{N}}^{\hat{a}_{s,\circlearrowright}|\hat{q}_{2}}$ for the CW input case become smaller or even zero within the same considered optical detuning (green dashed curve). These results imply that by setting distinct squeezing phase parameters for the counterpropagating optical modes, it could lead to asymmetric response in the entanglement generation between the coherent driving field and the mechanical mode. Moreover, it is clearly seen that the response asymmetry of $E_{\mathcal{N}}^{\hat{a}_{s,i}|\hat{q}_{j}}$ is dependent on both $\delta r_{i}$ and $\delta\theta_{i}$, with stronger asymmetry attainable through appropriate optimization of the squeezing phase condition. To clearly visualize this behavior, we define the asymmetric response ratio $\mathcal{I}_{j}$ as
\begin{align}
\mathcal{I}_{j}=\dfrac{E_{\mathcal{N}}^{\hat{a}_{s,\circlearrowleft}|\hat{q}_{j}}-E_{\mathcal{N}}^{\hat{a}_{s,\circlearrowright}|\hat{q}_{j}}}{E_{\mathcal{N}}^{\hat{a}_{s,\circlearrowleft}|\hat{q}_{j}}}.
\end{align}
As shown in Figs.\,\ref{Fig2}(b) and \ref{Fig2}(d), the behavior of $\mathcal{I}_{1}$ and $\mathcal{I}_{2}$ are similar with the variation of the squeezing parameters, which show greater sensitivity to $\delta r_{\circlearrowright}$ than to $\delta\theta_{\circlearrowright}$ and these isolation ratios could be more than $\textrm{0.9}$ when choosing $\delta r_{\circlearrowright}>\textrm{0.2}$ with proper values of $\delta\theta_{\circlearrowright}$. Besides, we have confirmed that the maximum values of $\mathcal{I}_{j}$ could even reach up to $1$ by properly choosing squeezing phase conditions, which means that such optomechanical entanglement can be unidirectionally produced only in one specific optical input direction when reversing the coherent driving field. Additionally, to investigate the robustness of such optomechanical entanglement against thermal noises, we further plot $E_{\mathcal{N}}^{\hat{\textit{a}}_{\textit{s},\textit{i}}|\hat{\textit{q}}_{\textit{1}}}$ and $E_{\mathcal{N}}^{\hat{\textit{a}}_{\textit{s},\textit{i}}|\hat{\textit{q}}_{\textit{2}}}$ for opposite optical input directions as functions of the mechanical environment temperature $T$ in Figs.\,\ref{Fig2}(e) and \ref{Fig2}(f). It is seen that in comparison with the CW input case, the entanglement with respect to the CCW input case is more robust against thermal noises, which can persist above a temperature of $\textrm{110}\,\textrm{mK}$.
\begin{figure*}[htbp]
\centering
\includegraphics[width=0.96\textwidth]{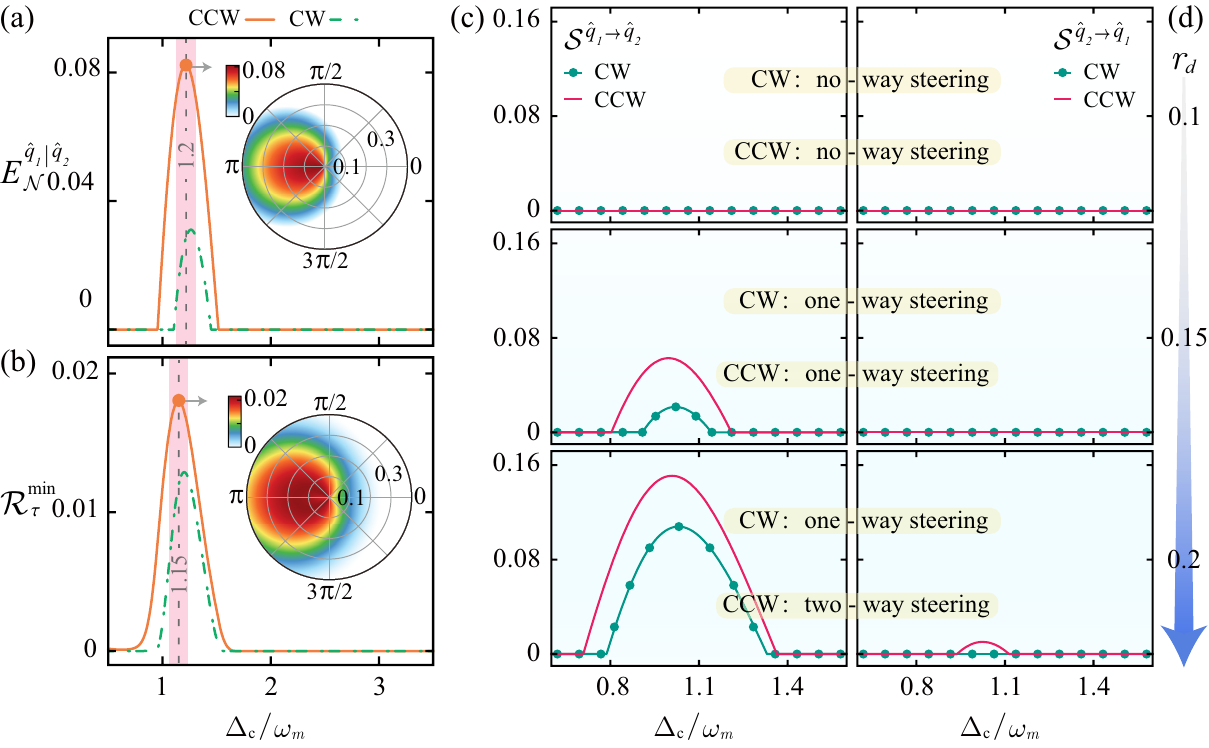}
\caption{\label{Fig4}Phase-selective generation of quantum entanglement and asymmetric EPR steering with respect to the two mechanical modes. (a),(b) The logarithmic negativity $E_{\mathcal{N}}^{\hat{q}_{1}|\hat{q}_{2}}$ and the minimum residual contangle $\mathcal{R}^{\textrm{min}}_{\tau}$ for opposite optical input directions are plotted as functions of the scaled optical detuning $\Delta_{c}/\omega_{m}$ with $r_{d}=0.2$ and $\theta_{d}=\pi$. The inserts show the dependence of $E_{\mathcal{N}}^{\hat{q}_{1}|\hat{q}_{2}}$ and $\mathcal{R}^{\textrm{min}}_{\tau}$ on the squeezing strength difference $\delta r_{\circlearrowright}$ and the squeezing phase difference $\delta\theta_{\circlearrowright}$ at the optical detuning $\Delta_{c}/\omega_{m}=1.2$ and $1.15$, respectively. (c),(d) The measure of quantum steerability $\mathcal{S}^{\hat{q}_{1}\rightarrow\hat{q}_{2}}$ and $\mathcal{S}^{\hat{q}_{2}\rightarrow\hat{q}_{1}}$ for opposite optical input directions are plotted as functions of the scaled optical detuning $\Delta_{c}/\omega_{m}$ under different squeezing strength $r_{d}$ with $\theta_{d}=\pi$. With the increase of $r_{d}$, EPR steering for the CCW input case can be stepwise driven from no-way regime, one-way regime to two-way regime, whereas for the CW input case, it can only be driven from no-way regime to one-way regime.}
\end{figure*}

Subsequently, we further investigate how to regulate the steerability of optomechanical entanglement and achieve one-way EPR steering through tuning the squeezing phase conditions. For this purpose, we plot the measure of quantum steerability $\mathcal{S}^{\hat{a}_{s,i}\rightarrow\hat{q}_{1}}$ and $\mathcal{S}^{\hat{q}_{1}\rightarrow\hat{a}_{s,i}}$ as functions of the scaled optical detuning $\Delta_{c}/\omega_{m}$ under different values of $r_{d}$ in Fig.\,\ref{Fig3}. It is seen that in terms of the CCW input case, for small values of $r_{d}$, both $\mathcal{S}^{\hat{a}_{s,\circlearrowleft}\rightarrow\hat{q}_{1}}$ and $\mathcal{S}^{\hat{q}_{1}\rightarrow\hat{a}_{s,\circlearrowleft}}$ are null within the considered optical detuning [see red solid lines in Figs.\,\ref{Fig3}(a) and \ref{Fig3}(b)], implying that the EPR steering are not present for both directions of $\hat{a}_{s,\circlearrowleft}\rightarrow\hat{q}_{1}$ and $\hat{q}_{1}\rightarrow\hat{a}_{s,\circlearrowleft}$. When increasing the values of the intracavity squeezing strength $r_{d}$, the quantum steerability $\mathcal{S}^{\hat{a}_{s,\circlearrowleft}\rightarrow\hat{q}_{1}}$ will increase monotonically and become nonzero [see Fig.\,\ref{Fig3}(a)], resulting in an asymmetric one-way EPR steering with $\mathcal{S}^{\hat{a}_{s,\circlearrowleft}\rightarrow\hat{q}_{1}}>0$ and $\mathcal{S}^{\hat{q}_{1}\rightarrow\hat{a}_{s,\circlearrowleft}}=0$. This is because quantum steerability is an increasing function of the effective COM coupling strength that can be enhanced by the intracavity squeezing strength $r_{d}$ [see Fig.\,\ref{Fig1}(c) for more details]. However, for the CW input case, since the optomechanical entanglement is absent, EPR steering, as a stronger correlated entangled state, can also not be produced, showing that $\mathcal{S}^{\hat{a}_{s,\circlearrowright}\rightarrow\hat{q}_{1}}$ and $\mathcal{S}^{\hat{q}_{1}\rightarrow\hat{a}_{s,\circlearrowright}}$ always keep zero with the variation of the controlling parameters [see green ball-solid lines in Figs.\,\ref{Fig3}(a) and \ref{Fig3}(b)].

By comparing the results of Figs.\,\ref{Fig2} and \ref{Fig3}, it can be intuitively found that when optomechanical entanglement is prepared in an asymmetric way, the generation of the corresponding EPR steering also exhibits asymmetric characteristics, with its directionality being asymmetrically controlled via intracavity squeezing parameters. Physically, the mechanism behind these phenomena can be understood as follows: As discussed previously in Fig.\,\ref{Fig1}(c), the optical input noises $N_{s,i}$ and $M_{s,i}$ can either be reduced or enhanced through adjusting the squeezing parameters $r_{e,i}$ and $\theta_{e,i}$, which provides an effective method to regulate the noise flows in CW and CCW modes in a controllable way. Here, as a specific example, we have chosen the parameter conditions where optical input noises are reduced for the CCW input case [corresponding to phase-matched condition] but amplified for the CW input case [corresponding to phase-mismatched condition]. This squeezing-phase-induced asymmetry could lead to differences in the quantum statistical properties of the CW and CCW modes. As shown in Fig.\,\ref{Fig1}(d), without the presence of coherent driving, the steady-state CM $V$ displays a dominant diagonal, which is consistent with a Gaussian distribution of a vacuum state. For the CCW input case, except for the dominant diagonal elements, the steady-state CM $V$ also displays some off-diagonal elements, which are not only present in the mechanical bipartition, but also emerges in the optomechanical bipartition. For the CW input case, due to the optical input noises $N_{s,\circlearrowright}\neq0$ and $M_{s,\circlearrowright}\neq0$, the off-diagonal elements with respect to the optomechanical bipartition vanish in steady-state CM $V$. These results indicate that the quantum correlations between the squeezed optical mode and the mechanical modes, enabled by the down-conversion interaction through nonlinear COM coupling, survive only with small thermal photon occupancy. Therefore, by asymmetrically choosing phase conditions for the CW and CCW modes, the phase-selective generation of optomechanical entanglement and EPR steering could be achieved in this configuration.

Finally, we further present how to achieve phase-selective generation of quantum entanglement and EPR steering involving two mechanical modes in Fig.\,\ref{Fig4}. Given that the mechanical entanglement is more fragile to thermal noises, the parameters for numerical simulation are altered as follows: $\kappa/\textrm{2}\pi=\textrm{14.4}\,\textrm{MHz}$, $\omega_{m}/\textrm{2}\pi=\textrm{16}\,\textrm{MHz}$, $Q_{m}\equiv\omega_{m}/\gamma_{m}=\textrm{10}$, $G_{1}/\textrm{2}\pi=\textrm{1.6}\,\textrm{MHz}$, $G_{2}/\textrm{2}\pi=\textrm{2.1}\,\textrm{MHz}$, $\lambda/\textrm{2}\pi=\textrm{4.8}\,\textrm{MHz}$, and $\bar{n}_{m}=\textrm{0.9}$. Figures\,\ref{Fig4}(a) and \ref{Fig4}(b) show the logarithmic negativity $E_{\mathcal{N}}^{\hat{q}_{1}|\hat{q}_{2}}$ and the minimum residual contangle $\mathcal{R}^{\textrm{min}}_{\tau}$ versus the scaled optical detuning $\Delta_{c}/\omega_{m}$ for opposite optical input directions. It is seen that for the CCW input case, by fulfilling the phase-matched condition, the bipartite mechanical entanglement and the tripartite entanglement are present within a finite interval of $\Delta_{c}$ around $\Delta_{c}/\omega_{m}\simeq\textrm{1.2}$. In contrast, for the CW input case, while the bipartite mechanical entanglement and the tripartite entanglement can still persist for certain values of mismatched phase (e.g., $\delta r_{\circlearrowright}=\textrm{0.4}$ and $\delta\theta_{\circlearrowright}=\pi$), their maximum values are suppressed compared to those in the CCW input case. Moreover, from the inserts of Figs.\,\ref{Fig4}(a) and \ref{Fig4}(b), it is also seen that when setting squeezing phase difference $\delta\theta_{\circlearrowright}=\textrm{0}$, the corresponding bipartite and tripartite entanglement can completely vanish at high values of $\delta r_{\circlearrowright}$. These results are consistent with the previous discussions for optomechanical entanglement, implying the achievement of phase-selective asymmetric response in bipartite mechanical entanglement and tripartite entanglement. In Figs.\,\ref{Fig4}(c) and \ref{Fig4}(d), to explore the behaviors of EPR steering between the two mechanical modes, we further plot the quantum steerability $\mathcal{S}^{\hat{q}_{1}\rightarrow\hat{q}_{2}}$ and $\mathcal{S}^{\hat{q}_{2}\rightarrow\hat{q}_{1}}$ versus the scaled optical detuning $\Delta_{c}/\omega_{m}$ under different squeezing strength $r_{d}$ with $\theta_{d}=\pi$. For the CCW input case, it is seen that accompanied with the occurrence of highly mechanical entangled state, we can get rich properties of EPR steering, where the overall state's asymmetry can be stepwise driven through the no-way regime ($\mathcal{S}^{\hat{q}_{1}\rightarrow\hat{q}_{2}}=\mathcal{S}^{\hat{q}_{2}\rightarrow\hat{q}_{1}}=0$), one-way regime ($\mathcal{S}^{\hat{q}_{2}\rightarrow\hat{q}_{1}}=0$ but $\mathcal{S}^{\hat{q}_{1}\rightarrow\hat{q}_{2}}>0$), and finally two-way regime ($\mathcal{S}^{\hat{q}_{1}\rightarrow\hat{q}_{2}}>0$ and $\mathcal{S}^{\hat{q}_{2}\rightarrow\hat{q}_{2}}>0$) with the increase of the intracavity squeezing strength $r_{d}$. However, for the CW input case, due to the phase-mismatch induced suppression of mechanical entanglement, it is seen that the directionality of EPR steering can only be driven from no-way regime to one-way regime under the same intracavity squeezing strength. From the above discussions, it is found that the mechanical EPR steering can be generated and manipulated in an asymmetric way. These results can be understood as follows: The proposed three-mode COM system has a bilinear cyclic coupling among the optical mode and the two mechanical modes. For an arbitrary bipartition $\mu$ and $\nu$, apart from their direct interaction, there is also an indirect interaction induced by their coupling to the third intermediate mode. The superposition of direct and indirect interaction paths can give rise to a quantum interference effect, whose strength is determined by their relative phase and coupling strength. Because both quantum entanglement and EPR steering increase monotonically with the coupling strength, their magnitudes can thus be modulated via the interference effect, resulting in the generation of tripartite entanglement. In this case, by regulating the two effective COM coupling strengths $\Lambda_{j}$ through tuning the intracavity squeezing $r_{d}$, the entanglement originally arising from nonlinear COM interactions could be transferred to the two mechanical modes via the interference effect. Moreover, since we have assumed different phase conditions for the CW and CCW modes, the bipartite and tripartite entanglement will be produced in an asymmetric way when the coherent driving field is applied from opposite directions. Also, due to the asymmetric COM coupling induced for the two mechanical modes, the overall asymmetry of EPR steering can be well controlled by the squeezing phase condition.

\section{Conclusion}\label{secIV}
In summary, we have presented how to achieve phase-selective generation and manipulation of quantum entanglement and asymmetric EPR steering in a three-mode squeezed COM system. Here, as a specific example, we have considered a near-field cavity COM setup to realize our proposal~\cite{Anetsberger2009NP,Sbarra2022NL}, which consists of a microdisk WGM resonator, a string of nanomechanical oscillators, and a tapered optical fiber. The WGM resonator supports two counterpropagating optical modes, which are degenerate and stay uncoupled. The coherent driving field from the waveguide is coupled into and out of the WGM resonator via evanescent coupling. Moreover, when placing the nanomechanical array close to the WGM resonator, it excites two mechanical modes due to the COM interaction induced by the WGM evanescent field. First, we show that when simultaneously applying two-photon pumping fields and broadband squeezed optical fields to CW and CCW modes, one can introduce a pair of squeezed optical modes, with each coupled to an independent squeezed vacuum reservoir. This configuration enables a phase-controlled light-reservoir interaction and allows one to tailor asymmetric noise flows for opposite input directions, which plays a key role in breaking the symmetry of the quantum statistical properties of the counterpropagating modes. Second, based on this unique feature of the squeezed COM system, it is found that the generation of various types of quantum entanglement and the associated EPR steering become phase-dependent and thus they can be produced in an asymmetric way. More interestingly, it is also found that by properly tuning the intracavity squeezing parameters, the directionality of EPR steering can be well controlled. According to this, we further show that with the increase of the squeezing strength, the directionality of EPR steering can be stepwise driven from no-way regime, one-way regime to two-way regime. Note that our proposal can also be realized with various types of multimode COM system, e.g., microwave electromechanical system~\cite{Kotler2021Science,Mercier2021Science}, cavity magnomechanical system~\cite{Shen2022PRL} , and nanospike-WGM optomechanical system~\cite{Pennetta2020PR}. These findings, opening up new perspectives for preparing rich types of quantum entanglement and EPR steering with asymmetric features, might be applicable for a variety of nascent quantum technologies ranging from one-sided device-independent quantum key
distribution to no-cloning quantum teleportation~\cite{Branciard2012PRA,Nathan2016Optica,Li2024PRL,He2015PRL2,Chiu2016NQI}.

\section{Acknowledgment}
H.J. is supported by the National Natural Science Foundation of China (NSFC, Grant No. 11935006, 12421005), the National Key R$\&$D Program of China (Grants No. 2024YFE0102400), and the Hunan Provincial Major Sci-Tech Program (Grant No. 2023ZJ1010). L.-M.K. is supported by the NSFC (Grants No. 12247105, 11935006, 12175060 and 12421005), the Hunan Provincial Major Sci-Tech Program (Grant No. 2023ZJ1010), and the Henan Science and Technology Major Project (Grant No. 241100210400). Y.-F.J. is supported by the NSFC (Grant No. 12405029) and the Natural Science Foundation of Henan Province (Grant No. 252300421221). D.-Y.W. is supported by the NSFC (Grant No. 12204424) and the China Postdoctoral Science Foundation (Grant No. 2022M722889). L.T. is supported by the NSFC (Grant No. 12305023) and the Sichuan Science and Technology Program (Grant No. 2024NSFSC1353). Y.W. is supported by the NSFC (Grant No. 12205256) and the Henan Provincial Science and Technology Research Project (Grant No. 232102221001). W.-S.B. is supported by the Henan Science and Technology Major Project of the Department of Science \& Technology of Henan Province (Grant No. 241100210400). Y.-L.Z. is supported by the Natural Science Foundation of Hunan Province (Grant No. 2025JJ60018) and the Scientific Research Foundation of Education Bureau of Hunan Province (Grant No.24B0866).

\appendix

\section{Derivation of the effective Hamiltonian}\label{AppA}
To diagonalize the optical mode of Hamiltonian\,(\ref{eq_Ham1}), one can perform a unitary Bogoliubov transformation with $U(\eta_{d})=\exp[(-\eta_{d}\hat{a}_{\circlearrowleft}^{\dagger2}+\eta_{d}^{\ast}\hat{a}_{\circlearrowleft}^{2})/2]$, where $\eta_{d}=r_{d}e^{-i\theta_{d}}$ is a complex squeezing parameter with $r_{d}$ the squeezing strength and $\theta_{d}$ the squeezing reference angle. By implementing this Bogoliubov transformation, a squeezed optical mode is introduced and the system Hamiltonian\,(\ref{eq_Ham1}) becomes
\begin{align}
\hat{\tilde{H}}_{\circlearrowleft}=&U^{\dagger}(\eta_{d})\hat{H}_{\circlearrowleft}U(\eta_{d}) \notag \\
=&\dfrac{\omega_{m,1}}{2}(\hat{p}_{1}^{2}+\hat{q}_{1}^{2})+\dfrac{\omega_{m,2}}{2}(\hat{p}_{2}^{2}+\hat{q}_{2}^{2})
+\lambda\hat{q}_{1}\hat{q}_{2}
\notag \\
&+\left[\Delta_{c}\cosh(2r_{d})-2\Xi_{d}\sinh(2r_{d})\right]\hat{a}_{s,\circlearrowleft}^{\dagger}\hat{a}_{s,\circlearrowleft}
\notag \\
&+\left[\Xi_{d}\cosh(2r_{d})-\dfrac{\Delta_{c}}{2}\sinh(2r_{d})\right]e^{-i\theta_{d}}\hat{a}_{s,\circlearrowleft}^{\dagger2}
\notag \\
&+\left[\Xi_{d}\cosh(2r_{d})-\dfrac{\Delta_{c}}{2}\sinh(2r_{d})\right]e^{i\theta_{d}}\hat{a}_{s,\circlearrowleft}^{2}
\notag \\
&-g_{1}\cosh(2r_{d})\hat{a}_{s,\circlearrowleft}^{\dagger}\hat{a}_{s,\circlearrowleft}\hat{q}_{1}-g_{2}\cosh(2r_{d})\hat{a}_{s,\circlearrowleft}^{\dagger}\hat{a}_{s,\circlearrowleft}\hat{q}_{2}
\notag \\
&+\dfrac{1}{2}g_{1}\sinh(2r_{d})(e^{-i\theta_{d}}\hat{a}_{s,\circlearrowleft}^{\dagger2}+e^{i\theta_{d}}\hat{a}_{s,\circlearrowleft}^{2})\hat{q}_{1}
\notag \\
&+\dfrac{1}{2}g_{2}\sinh(2r_{d})(e^{-i\theta_{d}}\hat{a}_{s,\circlearrowleft}^{\dagger2}+e^{i\theta_{d}}\hat{a}_{s,\circlearrowleft}^{2})\hat{q}_{2}
\notag \\
&-g_{1}\sinh^{2}(r_{d})\hat{q}_{1}-g_{2}\sinh^{2}(r_{d})\hat{q}_{2}
\notag \\
&+i\varepsilon_{d,\circlearrowleft}\cosh(r_{d})(\hat{a}_{s,\circlearrowleft}^{\dagger}-\hat{a}_{s,\circlearrowleft})
\notag \\
&+i\varepsilon_{d,\circlearrowleft}\sinh(r_{d})(e^{-i\theta_{d}}\hat{a}_{s,\circlearrowleft}^{\dagger}-e^{i\theta_{d}}\hat{a}_{s,\circlearrowleft})
\notag \\
&+\Delta_{c}\sinh^{2}(r_{d})-\Xi_{d}\sinh(2r_{d}).
\end{align}
By setting the coefficients of the quadratic terms $\hat{a}_{s,\circlearrowleft}^{\dagger2}$ and $\hat{a}_{s,\circlearrowleft}^{2}$ to be zero, i.e.,
\begin{align}
\Delta_{c}\cosh(2r_{d})-2\Xi_{d}\sinh(2r_{d})=0,
\end{align}
we have $r_{d}=(1/4)\ln[(\Delta_{c}+2\Xi_{d})/(\Delta_{c}-2\Xi_{d})]$. Correspondingly, the effective Hamiltonian of the system is obtained as
\begin{align}
\hat{\tilde{H}}_{\circlearrowleft}=&\omega_{s}\hat{a}_{s,\circlearrowleft}^{\dagger}\hat{a}_{s,\circlearrowleft}+\dfrac{\omega_{m,1}}{2}(\hat{p}_{1}^{2}+\hat{q}_{1}^{2})+\dfrac{\omega_{m,2}}{2}(\hat{p}_{2}^{2}+\hat{q}_{2}^{2})
\notag \\
&-\zeta_{s,1}\hat{a}_{s,\circlearrowleft}^{\dagger}\hat{a}_{s,\circlearrowleft}\hat{q}_{1}+\dfrac{\zeta_{p,1}}{2}(e^{-i\theta_{d}}\hat{a}_{s,\circlearrowleft}^{\dagger2}+e^{i\theta_{d}}\hat{a}_{s,\circlearrowleft}^{2})\hat{q}_{1}
\notag \\
&-\zeta_{s,2}\hat{a}_{s,\circlearrowleft}^{\dagger}\hat{a}_{s,\circlearrowleft}\hat{q}_{2}+\dfrac{\zeta_{p,2}}{2}(e^{-i\theta_{d}}\hat{a}_{s,\circlearrowleft}^{\dagger2}+e^{i\theta_{d}}\hat{a}_{s,\circlearrowleft}^{2})\hat{q}_{2}
\notag \\
&\!-\!F_{1}\hat{q}_{1}\!-\!F_{2}\hat{q}_{2}\!+\!\lambda\hat{q}_{1}\hat{q}_{2}
\!+\!i\varepsilon_{d,\circlearrowleft}\!\cosh(r_{d})(\hat{a}_{s,\circlearrowleft}^{\dagger}\!-\!\hat{a}_{s,\circlearrowleft})
\notag \\
&+i\varepsilon_{d,\circlearrowleft}\sinh(r_{d})(e^{-i\theta_{d}}\hat{a}_{s,\circlearrowleft}^{\dagger}-e^{i\theta_{d}}\hat{a}_{s,\circlearrowleft})+C,
\label{eq_H2}
\end{align}
where
\begin{align}
\nonumber
\omega_{s}\!&=\!\Delta_{c}\cosh(2r_{d})\!-\!2\Xi_{d}\sinh(2r_{d})\!=\!(\Delta_{c}\!-\!2\Xi_{d})\exp(2r_{d}),\\ \nonumber
\zeta_{s,j}&=\dfrac{g_{j}\Delta_{c}}{\sqrt{\Delta_{c}^{2}-4\Xi_{d}^{2}}}=g_{j}\cosh(2r_{d}),\notag\\
\zeta_{p,j}&=\dfrac{2g_{j}\Xi_{d}}{\sqrt{\Delta_{c}^{2}-4\Xi_{d}^{2}}}=g_{j}\sinh(2r_{d}),\notag\\
F_{j}&=g_{j}\sinh^{2}(r_{d}),~C=\Delta_{c}\sinh^{2}(r_{d})-\Xi_{d}\sinh(2r_{d}).
\end{align}

\section{Derivation of the effective master equation under phase-matched condition}\label{AppB}
To regulate the light-reservoir interaction, we have considered the injection of a broadband squeezed optical field, which can act as a squeezed-vacuum reservoir. We assume the squeezed optical field, with squeezing strength $ r_{e} $ and reference phase angle $ \theta_{e} $, is around the central frequency $ \omega_{c} $. Besides, the two mechanical modes are assumed to be coupled with two independent thermal reservoirs at the same bath temperature $T$. Then, including the dissipations of the optical and mechanical modes, the dynamics of the total system in the original picture is governed by the Born-Markovian master equation~\cite{Zoller2000book}
\begin{align}
\dot{\rho}_{\circlearrowleft}=&i\big[\rho_{\circlearrowleft},\hat{H}_{\circlearrowleft}\big]+\frac{\kappa}{2}(N_{e,\circlearrowleft}+1)\mathcal{D}[\hat{a}_{\circlearrowleft}]\rho_{\circlearrowleft}+\frac{\kappa}{2}N_{e,\circlearrowleft}\mathcal{D}[\hat{a}_{\circlearrowleft}^{\dagger}]\rho_{\circlearrowleft}
\notag \\
&-\frac{\kappa}{2}M_{e,\circlearrowleft}\mathcal{G}[\hat{a}_{\circlearrowleft}]\rho_{\circlearrowleft}-\frac{\kappa}{2}M_{e,\circlearrowleft}^{\ast}\mathcal{G}[\hat{a}_{\circlearrowleft}^{\dagger}]\rho_{\circlearrowleft}
\notag \\
&-\sum_{j=1,2}\left(i\dfrac{\gamma_{m,j}}{2}[\hat{q}_{j},\{\hat{p}_{j},\rho_{\circlearrowleft}\}]+\gamma_{m,j}\bar{n}_{m,j}[\hat{q}_{j},[\hat{q}_{j},\rho_{\circlearrowleft}]]\right), \end{align}
where
\begin{align}
\mathcal{D}[\hat{a}_{\circlearrowleft}]\rho_{\circlearrowleft}&=2\hat{a}_{\circlearrowleft}\rho_{\circlearrowleft}\hat{a}_{\circlearrowleft}^{\dagger}-(\hat{a}_{\circlearrowleft}^{\dagger}\hat{a}_{\circlearrowleft}\rho_{\circlearrowleft}+\rho_{\circlearrowleft}\hat{a}_{\circlearrowleft}^{\dagger}\hat{a}_{\circlearrowleft}),
\notag \\
\mathcal{G}[\hat{a}_{\circlearrowleft}]\rho_{\circlearrowleft}&=2\hat{a}_{\circlearrowleft}\rho_{\circlearrowleft}\hat{a}_{\circlearrowleft}-(\hat{a}_{\circlearrowleft}\hat{a}_{\circlearrowleft}\rho_{\circlearrowleft}+\rho_{\circlearrowleft}\hat{a}_{\circlearrowleft}\hat{a}_{\circlearrowleft}),
\end{align}
and $\kappa~(\gamma_{m,j})$ is the optical (mechanical) decay rate. $\bar{n}_{m,j}=1/[\exp(\omega_{m,j}/k_{B}T)-1]$ is the thermal phonon number, while $N_{e,\circlearrowleft}=\sinh^{2}(r_{e})$ and $M_{e,\circlearrowleft}=\cosh(r_{e})\sinh(r_{e})e^{i\theta_{e}}$ describe the dissipation and the two-photon correlation of the cavity field caused by the squeezed-vacuum reservoir, respectively.

By introducing the Bogoliubov transformation in Eq.\,(\ref{eq_bogo}), the Lindblad operators becomes
\begin{align}
&U^{\dagger}(\eta_{d})\mathcal{D}[\hat{a}_{\circlearrowleft}]\rho_{\circlearrowleft}U(\eta_{d}) \notag \\
=&2\cosh^{2}(r_{d})\hat{a}_{s,\circlearrowleft}\tilde{\rho}_{\circlearrowleft}\hat{a}_{s,\circlearrowleft}^{\dagger}+2\sinh^{2}(r_{d})\hat{a}_{s,\circlearrowleft}^{\dagger}\tilde{\rho}_{\circlearrowleft}\hat{a}_{s,\circlearrowleft}
\notag \\
&-\sinh(2r_{d})(e^{-i\theta_{d}}\hat{a}_{s,\circlearrowleft}^{\dagger}\tilde{\rho}_{\circlearrowleft}\hat{a}_{s,\circlearrowleft}^{\dagger}+e^{i\theta_{d}}\hat{a}_{s,\circlearrowleft}\tilde{\rho}_{\circlearrowleft}\hat{a}_{s,\circlearrowleft})
\notag \\
&-\left[\cosh^{2}(r_{d})\hat{a}_{s,\circlearrowleft}^{\dagger}\hat{a}_{s,\circlearrowleft}\tilde{\rho}_{\circlearrowleft}+\sinh^{2}(r_{d})\hat{a}_{s,\circlearrowleft}\hat{a}_{s,\circlearrowleft}^{\dagger}\tilde{\rho}_{\circlearrowleft}\right.
\notag \\
&\left.-\sinh(r_{d})\cosh(r_{d})(e^{-i\theta_{d}}\hat{a}_{s,\circlearrowleft}^{\dagger}\hat{a}_{s,\circlearrowleft}^{\dagger}\tilde{\rho}_{\circlearrowleft}+e^{i\theta_{d}}\hat{a}_{s,\circlearrowleft}\hat{a}_{s,\circlearrowleft}\tilde{\rho}_{\circlearrowleft})\right]
\notag \\
&-\left[\cosh^{2}(r_{d})\tilde{\rho}_{\circlearrowleft}\hat{a}_{s,\circlearrowleft}^{\dagger}\hat{a}_{s,\circlearrowleft}+\sinh^{2}(r_{d})\tilde{\rho}_{\circlearrowleft}\hat{a}_{s,\circlearrowleft}\hat{a}_{s,\circlearrowleft}^{\dagger}\right.
\notag \\
&\left.-\sinh(r_{d})\cosh(r_{d})(e^{-i\theta_{d}}\tilde{\rho}_{\circlearrowleft}\hat{a}_{s,\circlearrowleft}^{\dagger}\hat{a}_{s,\circlearrowleft}^{\dagger}+e^{i\theta_{d}}\tilde{\rho}_{\circlearrowleft}\hat{a}_{s,\circlearrowleft}\hat{a}_{s,\circlearrowleft})\right],
\notag
\end{align}

\begin{align}
&U^{\dagger}(\eta_{d})\mathcal{G}[\hat{a}_{\circlearrowleft}]\rho_{\circlearrowleft}U(\eta_{d}) \notag \\
=&2e^{-2i\theta_{d}}\sinh^{2}(r_{d})\hat{a}_{s,\circlearrowleft}^{\dagger}\tilde{\rho}_{\circlearrowleft}\hat{a}_{s,\circlearrowleft}^{\dagger}+2\cosh^{2}(r_{d})\hat{a}_{s,\circlearrowleft}\tilde{\rho}_{\circlearrowleft}\hat{a}_{s,\circlearrowleft}
\notag \\
&-e^{-i\theta_{d}}\sinh(2r_{d})(\hat{a}_{s,\circlearrowleft}^{\dagger}\tilde{\rho}_{\circlearrowleft}\hat{a}_{s,\circlearrowleft}+\hat{a}_{s,\circlearrowleft}\tilde{\rho}_{\circlearrowleft}\hat{a}_{s,\circlearrowleft}^{\dagger})
\notag \\
&-\left[e^{-2i\theta_{d}}\sinh^{2}(r_{d})\hat{a}_{s,\circlearrowleft}^{\dagger}\hat{a}_{s,\circlearrowleft}^{\dagger}\tilde{\rho}_{\circlearrowleft}+\cosh^{2}(r_{d})\hat{a}_{s,\circlearrowleft}\hat{a}_{s,\circlearrowleft}\tilde{\rho}_{\circlearrowleft}\right.
\notag \\
&\left.-e^{-i\theta_{d}}\sinh(r_{d})\cosh(r_{d})(\hat{a}_{s,\circlearrowleft}^{\dagger}\hat{a}_{s,\circlearrowleft}\tilde{\rho}_{\circlearrowleft}+\hat{a}_{s,\circlearrowleft}\hat{a}_{s,\circlearrowleft}^{\dagger}\tilde{\rho}_{\circlearrowleft})\right]
\notag \\
&-\left[e^{-2i\theta_{d}}\sinh^{2}(r_{d})\tilde{\rho}_{\circlearrowleft}\hat{a}_{s,\circlearrowleft}^{\dagger}\hat{a}_{s,\circlearrowleft}^{\dagger}+\cosh^{2}(r_{d})\tilde{\rho}_{\circlearrowleft}\hat{a}_{s,\circlearrowleft}\hat{a}_{s,\circlearrowleft}\right.
\notag \\
&\left.-e^{-i\theta_{d}}\sinh(r_{d})\cosh(r_{d})(\tilde{\rho}_{\circlearrowleft}\hat{a}_{s,\circlearrowleft}^{\dagger}\hat{a}_{s,\circlearrowleft}+\tilde{\rho}_{\circlearrowleft}\hat{a}_{s,\circlearrowleft}\hat{a}_{s,\circlearrowleft}^{\dagger})\right].
\end{align}
In this case, the effective Born-Markovian master equation could be rewritten as
\begin{align}
\dot{\tilde{\rho}}_{\circlearrowleft}=&U^{\dagger}(\eta_{d})\dot{\rho}_{\circlearrowleft}U(\eta_{d})
\notag \\
=&i\big[\tilde{\rho}_{\circlearrowleft},\hat{\tilde{H}}_{\circlearrowleft}\big]+\frac{\kappa}{2}(N_{s,\circlearrowleft}+1)\mathcal{D}[\hat{a}_{s,\circlearrowleft}]\tilde{\rho}_{\circlearrowleft}
+\frac{\kappa}{2}N_{s,\circlearrowleft}\mathcal{D}[\hat{a}_{s,\circlearrowleft}^{\dagger}]\tilde{\rho}_{\circlearrowleft}
\notag \\
&-\frac{\kappa}{2}M_{s,\circlearrowleft}\mathcal{G}[\hat{a}_{s,\circlearrowleft}]\tilde{\rho}_{\circlearrowleft}
-\frac{\kappa}{2}M_{s,\circlearrowleft}^{\ast}\mathcal{G}[\hat{a}_{s,\circlearrowleft}]\tilde{\rho}_{\circlearrowleft}
\notag \\
&-\sum_{j=1,2}\left(i\dfrac{\gamma_{m,j}}{2}[\hat{q}_{j},\{\hat{p}_{j},\tilde{\rho}_{\circlearrowleft}\}]+\gamma_{m,j}\bar{n}_{m,j}[\hat{q}_{j},[\hat{q}_{j},\tilde{\rho}_{\circlearrowleft}]]\right),
\end{align}
where
\begin{align}
\nonumber
N_{s,\circlearrowleft}=&\sinh^{2}(r_{d})\cosh^{2}(r_{e,\circlearrowleft})+\cosh^{2}(r_{d})\sinh^{2}(r_{e,\circlearrowleft})
\notag \\
&+\dfrac{1}{2}\cos\delta\theta\sinh(2r_{d})\sinh(2r_{e,\circlearrowleft}),
\notag\\
M_{s,\circlearrowleft}\!=&e^{i\theta_{d}}\![\sinh(r_{d})\!\cosh(r_{e,\circlearrowleft})\!+\!e^{i\delta\theta_{\circlearrowleft}}\!\cosh(r_{d})\!\sinh(r_{e,\circlearrowleft})\!]
\notag\\
&\times\![\cosh(r_{d})\!\cosh(r_{e,\circlearrowleft})\!+\!e^{-i\delta\theta_{\circlearrowleft}}\!\sinh(r_{d})\!\sinh(r_{e,\circlearrowleft})].
\end{align}
Here $N_{s,\circlearrowleft}$ and $M_{s,\circlearrowleft}$ denote the effective thermal noise and two-photon-correlation strength, respectively, which can be simplified in case of $r_{d}=r_{e,\circlearrowleft}=r$, i.e.,
\begin{align}
N_{s,\circlearrowleft}=&\dfrac{1}{2}\sinh^{2}(2r)[1+\cos(\theta_{e,\circlearrowleft}-\theta_{d})],
\notag\\
M_{s,\circlearrowleft}=&\dfrac{1}{2}e^{i\theta_{d}}\sinh(2r)[1+e^{i(\theta_{e,\circlearrowleft}-\theta_{d})}]
\notag\\
&\times[\cosh^{2}(r)+e^{-i(\theta_{e,\circlearrowleft}-\theta_{d})}\sinh^{2}(r)].
\end{align}
Obviously, the thermal noise $N_{s,\circlearrowleft}$ and the two-photon-correlation strength $M_{s,\circlearrowleft}$ can be completely eliminated by choosing $ r_{d}=r_{e,\circlearrowleft} $ and $\theta_{e,\circlearrowleft}-\theta_{d}=\pm n\pi (n=1,3,5...)$, i.e., $N_{s,\circlearrowleft}=M_{s,\circlearrowleft}=0$. The reservoir of the original optical mode is squeezed along the axis with angle $\dfrac{\theta_{e,\circlearrowleft}}{2}$. In the basis of the squeezed optical mode $\hat{a}_{s,\circlearrowleft}$, this effect can be canceled by the squeezing (along axis $\dfrac{\theta_{d}}{2}$) induced by the parametric amplification of $\hat{a}_{\circlearrowleft}$, when $r_{d}=r_{e,\circlearrowleft}$ and $\theta_{e,\circlearrowleft}+\theta_{d}=\pm n\pi (n=1,3,5...)$, yielding the phase-matched condition. This means the squeezed-vacuum reservoir of $\hat{a}_{\circlearrowleft}$ corresponds to an effective vacuum reservoir of $\hat{a}_{s,\circlearrowleft}$ under the phase-matched condition. In this case, the effective Born-Markovian master equation is derived as
\begin{align}
\dot{\tilde{\rho}}_{\circlearrowleft}
\!=&i\!\left[\tilde{\rho}_{\circlearrowleft},\hat{\tilde{H}}_{\circlearrowleft}\right]
\!+\!\frac{\kappa}{2}\mathcal{D}[\hat{a}_{s,\circlearrowleft}]\tilde{\rho}_{\circlearrowleft}
\!-\!\sum_{j=1,2}\left(i\dfrac{\gamma_{m,j}}{2}[\hat{q}_{j},\{\hat{p}_{j},\tilde{\rho}_{\circlearrowleft}\}]\right.
\notag \\
&\left.+\gamma_{m,j}\bar{n}_{m,j}[\hat{q}_{j},[\hat{q}_{j},\tilde{\rho}_{\circlearrowleft}]]\right),
\end{align}
where
\begin{align}
\mathcal{D}[\hat{a}_{s,\circlearrowleft}]\tilde{\rho}_{\circlearrowleft}=2\hat{a}_{s,\circlearrowleft}\tilde{\rho}_{\circlearrowleft}\hat{a}_{s,\circlearrowleft}^{\dagger}-(\hat{a}_{s,\circlearrowleft}^{\dagger}\hat{a}_{s,\circlearrowleft}\tilde{\rho}_{\circlearrowleft}+\tilde{\rho}_{\circlearrowleft}\hat{a}_{s,\circlearrowleft}^{\dagger}\hat{a}_{s,\circlearrowleft})
\end{align}
are the effective Lindblad operators, while $[\cdot,\cdot]$ and $\{\cdot,\cdot\}$ denote the commutator and anti-commutator, respectively.

\section{Derivation of the dynamics of the quantum correlation function}\label{AppC}
By employing the linearized effective master equation (\ref{eq_meq3}), the dynamics of the quantum correlation function in the vector $X$ can be derived as
\begin{align}
\dot{x}_{1}
=&-\kappa x_{1}
+i\Lambda_{1}x_{14}
-i\Lambda_{1}^{\ast}x_{13}
+i\Lambda_{2}x_{18}
\notag\\
&-i\Lambda_{2}^{\ast}x_{17}
+\kappa N_{s},
\end{align}

\begin{align}
\dot{x}_{2}
=&-\kappa x_{2}
+i\Lambda_{1}x_{14}
-i\Lambda_{1}^{\ast}x_{13}
+i\Lambda_{2}x_{18}
\notag\\
&-i\Lambda_{2}^{\ast}x_{17}
+\kappa(N_{s}+1),
\end{align}

\begin{align}
\dot{x}_{3}
=\omega_{m,1}x_{9}
+\omega_{m,1}x_{10},
\end{align}

\begin{align}
\dot{x}_{4}
=&-2\gamma_{m,1}x_{4}
-\omega_{m,1}x_{9}
-\omega_{m,1}x_{10}
+2\Lambda_{1}x_{16}
\notag\\
&+2\Lambda_{1}^{\ast}x_{15}
-2\lambda x_{24}
+2\gamma_{m,1}\bar{n}_{m,1},
\end{align}

\begin{align}
\dot{x}_{5}
=\omega_{m,2}x_{11}
+\omega_{m,2}x_{12},
\end{align}

\begin{align}
\dot{x}_{6}
=&-2\gamma_{m,2}x_{6}
-\omega_{m,2}x_{11}
-\omega_{m,2}x_{12}
+2\Lambda_{2}x_{20}
\notag\\
&+2\Lambda_{2}^{\ast}x_{19}
-2\lambda x_{23}
+2\gamma_{m,2}\bar{n}_{m,2},
\end{align}

\begin{align}
\dot{x}_{7}
=-(2i\Delta_{s}+\kappa)x_{7}
+2i\Lambda_{1}x_{13}
+2i\Lambda_{2}x_{17}
+\kappa M_{s}^{\ast},
\end{align}

\begin{align}
\dot{x}_{8}
=(2i\Delta_{s}-\kappa)x_{8}
-2i\Lambda_{1}^{\ast}x_{14}
-2i\Lambda_{2}^{\ast}x_{18}
+\kappa M_{s},
\end{align}

\begin{align}
\dot{x}_{9}
=&-\dfrac{\gamma_{m,1}}{2}x_{9}
-\dfrac{\gamma_{m,1}}{2}x_{10}
+\omega_{m,1}x_{4}
-\omega_{m,1}x_{3}
\notag\\
&-\lambda x_{21}
+\Lambda_{1}x_{14}
+\Lambda_{1}^{\ast}x_{13},
\end{align}

\begin{align}
\dot{x}_{10}
=&-\dfrac{\gamma_{m,1}}{2}x_{10}
-\dfrac{\gamma_{m,1}}{2}x_{9}
+\omega_{m,1}x_{4}
-\omega_{m,1}x_{3}
\notag\\
&-\lambda x_{21}
+\Lambda_{1}x_{14}
+\Lambda_{1}^{\ast}x_{13},
\end{align}

\begin{align}
\dot{x}_{11}
=&-\dfrac{\gamma_{m,2}}{2}x_{11}
-\dfrac{\gamma_{m,2}}{2}x_{12}
+\omega_{m,2}x_{6}
-\omega_{m,2}x_{5}
\notag\\
&-\lambda x_{21}
+\Lambda_{2}x_{18}
+\Lambda_{2}^{\ast}x_{17},
\end{align}

\begin{align}
\dot{x}_{12}
=&-\dfrac{\gamma_{m,2}}{2}x_{12}
-\dfrac{\gamma_{m,2}}{2}x_{11}
+\omega_{m,2}x_{6}
-\omega_{m,2}x_{5}
\notag\\
&-\lambda x_{21}
+\Lambda_{2}x_{18}
+\Lambda_{2}^{\ast}x_{17},
\end{align}

\begin{align}
\dot{x}_{13}
=&-\left(i\Delta_{s}+\dfrac{\kappa}{2}\right)x_{13}
+\omega_{m,1}x_{15}
\notag\\
&+i\Lambda_{1}x_{3}
+i\Lambda_{2}x_{21},
\end{align}

\begin{align}
\dot{x}_{14}
=&\left(i\Delta_{s}-\dfrac{\kappa}{2}\right)x_{14}
+\omega_{m,1}x_{16}
\notag\\
&-i\Lambda_{1}^{\ast}x_{3}
-i\Lambda_{2}^{\ast}x_{21},
\end{align}

\begin{align}
\dot{x}_{15}
=&-\left(i\Delta_{s}+\gamma_{m,1}+\dfrac{\kappa}{2}\right)x_{15}
-\omega_{m,1}x_{13}
+\Lambda_{1}^{\ast}x_{7}
\notag\\
&+\Lambda_{1}x_{2}
+i\Lambda_{1}x_{9}
+i\Lambda_{2}x_{24}
-\lambda x_{17},
\end{align}

\begin{align}
\dot{x}_{16}
=&\left(i\Delta_{s}-\gamma_{m,1}-\dfrac{\kappa}{2}\right)x_{16}
-\omega_{m,1}x_{14}
+\Lambda_{1}x_{8}
\notag\\
&+\Lambda_{1}^{\ast}x_{1}
-i\Lambda_{1}^{\ast}x_{9}
-i\Lambda_{2}^{\ast}x_{24}
-\lambda x_{18},
\end{align}

\begin{align}
\dot{x}_{17}
=&-\left(i\Delta_{s}+\dfrac{\kappa}{2}\right)x_{17}
+\omega_{m,1}x_{19}
\notag\\
&+i\Lambda_{1}x_{21}
+i\Lambda_{2}x_{5},
\end{align}

\begin{align}
\dot{x}_{18}
=&\left(i\Delta_{s}-\dfrac{\kappa}{2}\right)x_{18}
+\omega_{m,1}x_{20}
\notag\\
&-i\Lambda_{1}^{\ast}x_{21}
-i\Lambda_{2}^{\ast}x_{5},
\end{align}

\begin{align}
\dot{x}_{19}
=&-\left(i\Delta_{s}+\gamma_{m,2}+\dfrac{\kappa}{2}\right)x_{19}
-\omega_{m,2}x_{17}
+i\Lambda_{1}x_{23}
\notag\\
&+\Lambda_{2}x_{2}
+\Lambda_{2}^{\ast}x_{7}
+i\Lambda_{2}x_{11}
-\lambda x_{13},
\end{align}

\begin{align}
\dot{x}_{20}
=&\left(i\Delta_{s}-\gamma_{m,2}-\dfrac{\kappa}{2}\right)x_{20}
-\omega_{m,2}x_{18}
-i\Lambda_{1}^{\ast}x_{23}
\notag\\
&+\Lambda_{2}^{\ast}x_{1}
+\Lambda_{2}x_{8}
-i\Lambda_{2}^{\ast}x_{11}
-\lambda x_{14},
\end{align}

\begin{align}
\dot{x}_{21}
=\omega_{m,1}x_{24}
+\omega_{m,2}x_{23},
\end{align}

\begin{align}
\dot{x}_{22}
=&-(\gamma_{m,1}+\gamma_{m,2})x_{22}
-\omega_{m,1}x_{23}
-\omega_{m,2}x_{24}
-\lambda x_{10}
\notag \\
&-\lambda x_{11}
+\Lambda_{1}x_{20}
+\Lambda_{1}^{\ast}x_{19}
+\Lambda_{2}x_{16}
+\Lambda_{2}^{\ast}x_{15},
\end{align}

\begin{align}
\dot{x}_{23}
=&-\gamma_{m,2}x_{23}
+\omega_{m,1}x_{22}
-\omega_{m,2}x_{21}
\notag\\
&-\lambda x_{3}
+\Lambda_{2}x_{14}
+\Lambda_{2}^{\ast}x_{13},
\end{align}

\begin{align}
\dot{x}_{24}
=&-\gamma_{m,1}x_{24}
-\omega_{m,1}x_{21}
+\omega_{m,2}x_{22}
\notag\\
&-\lambda x_{5}
+\Lambda_{1}x_{18}
+\Lambda_{1}^{\ast}x_{17}.
\end{align}
By rewriting the above equations in a compact form, i.e.,
\begin{align}
\dot{X}=A\cdot X+b,
\end{align}
and introducing the following expressions
\begin{align}
\Omega_{1,\pm}&=i\Delta_{s}\pm\dfrac{\kappa}{2}, \notag \\
\Omega_{2,\pm}&=i\Delta_{s}\pm\gamma_{m,1}\pm\dfrac{\kappa}{2}, \notag \\
\Omega_{3,\pm}&=i\Delta_{s}\pm\gamma_{m,2}\pm\dfrac{\kappa}{2}, \notag \\
\Omega_{4}&=\gamma_{m,1}+\gamma_{m,2}.
\end{align}
we have
\begin{align}
A=\left(\begin{matrix}
A_{11} & A_{12} & A_{13} & A_{14}\\
A_{21} & A_{22} & A_{23} & A_{24}\\
A_{31} & A_{32} & A_{33} & A_{34}\\
A_{41} & A_{42} & A_{43} & A_{44}\\
\end{matrix}\right),
\end{align}
\begin{align}
b=&\left[\kappa N_{s},\kappa(N_{s}+1),0,2\gamma_{m,1}\bar{n}_{m,1},0,2\gamma_{m,2}\bar{n}_{m,2},\kappa M_{s}^{\ast},\right.
\notag \\
&\left.\kappa M_{s},0,0,0,0,0,0,0,0,0,0,0,0,0,0,0,0\right]^{T},
\end{align}
and
\renewcommand{\arraystretch}{1}
\begin{align}
A_{11}=\left(\begin{matrix}
-\kappa & 0 & 0 & 0 & 0 & 0 \\
0 &-\kappa & 0 & 0 & 0 & 0 \\
0 & 0 & 0 & 0 & 0 & 0 \\
0 & 0 & 0 & -2\gamma_{m,1} & 0 & 0 \\
0 & 0 & 0 & 0 & 0 & 0 \\
0 & 0 & 0 & 0 & 0 & -2\gamma_{m,2}
\end{matrix}\right)\notag,
\end{align}

\begin{align}
A_{12}=\left(\begin{matrix}
0 & 0 & 0 & 0 & 0 & 0 \\
0 & 0 & 0 & 0 & 0 & 0 \\
0 & 0 & \omega_{m,1} & \omega_{m,1} & 0 & 0 \\
0 & 0 & -\omega_{m,1} & -\omega_{m,1} & 0 & 0 \\
0 & 0 & 0 & 0 & \omega_{m,2} & \omega_{m,2} \\
0 & 0 & 0 & 0 & -\omega_{m,2} & -\omega_{m,2}
\end{matrix}\right)\notag,
\end{align}

\begin{align}
A_{13}=\left(\begin{matrix}
-i\Lambda_{1}^{\ast} & i\Lambda_{1} & 0 & 0 & -i\Lambda_{2}^{\ast} & i\Lambda_{2} \\
-i\Lambda_{1}^{\ast} & i\Lambda_{1} & 0 & 0 & -i\Lambda_{2}^{\ast} & i\Lambda_{2} \\
0 & 0 & 0 & 0 & 0 & 0 \\
0 & 0 & 2\Lambda_{1}^{\ast} & 2\Lambda_{1} & 0 & 0 \\
0 & 0 & 0 & 0 & 0 & 0 \\
0 & 0 & 0 & 0 & 0 & 0
\end{matrix}\right)\notag,
\end{align}

\begin{align}
A_{14}=\left(\begin{matrix}
0 & 0 & 0 & 0 & 0 & 0 \\
0 & 0 & 0 & 0 & 0 & 0\\
0 & 0 & 0 & 0 & 0 & 0 \\
0 & 0 & 0 & 0 & 0 & -2\lambda\\
0 & 0 & 0 & 0 & 0 & 0 \\
2\Lambda_{2}^{\ast} & 2\Lambda_{2} & 0 & 0 & -2\lambda & 0
\end{matrix}\right)\notag,
\end{align}

\begin{align}
A_{21}=\left(\begin{matrix}
0 & 0 & 0 & 0 & 0 & 0 \\
0 & 0 & 0 & 0 & 0 & 0 \\
0 & 0 & -\omega_{m,1} & \omega_{m,1} & 0 & 0 \\
0 & 0 & -\omega_{m,1} & \omega_{m,1} & 0 & 0 \\
0 & 0 & 0 & 0 & -\omega_{m,2} & \omega_{m,2} \\
0 & 0 & 0 & 0 & -\omega_{m,2} & \omega_{m,2}
\end{matrix}\right)\notag,
\end{align}

\begin{align}
\setlength{\arraycolsep}{0.5pt}
A_{22}=\dfrac{1}{2}\left(\begin{matrix}
-4\Omega_{1,+} & 0 & 0 & 0 & 0 & 0 \\
0 & 4\Omega_{1,-} & 0 & 0 & 0 & 0 \\
0 & 0 & -\gamma_{m,1} & -\gamma_{m,1} & 0 & 0 \\
0 & 0 & -\gamma_{m,1} & -\gamma_{m,1} & 0 & 0 \\
0 & 0 & 0 & 0 & -\gamma_{m,2} & -\gamma_{m,2} \\
0 & 0 & 0 & 0 & -\gamma_{m,2} & -\gamma_{m,2}
\end{matrix}\right)\notag,
\end{align}

\begin{align}
A_{23}=\left(\begin{matrix}
2i\Lambda_{1} & 0 & 0 & 0 & 2i\Lambda_{2} & 0 \\
0 & -2i\Lambda_{1}^{\ast} & 0 & 0 & 0 & -2i\Lambda_{2}^{\ast} \\
\Lambda_{1}^{\ast} & \Lambda_{1} & 0 & 0 & 0 & 0 \\
\Lambda_{1}^{\ast} & \Lambda_{1} & 0 & 0 & 0 & 0 \\
0 & 0 & 0 & 0 & \Lambda_{2}^{\ast} & \Lambda_{2} \\
0 & 0 & 0 & 0 & \Lambda_{2}^{\ast} & \Lambda_{2}
\end{matrix}\right)\notag,
\end{align}

\begin{align}
A_{24}=\left(\begin{matrix}
0 & 0 & 0 & 0 & 0 & 0 \\
0 & 0 & 0 & 0 & 0 & 0 \\
0 & 0 & -\lambda & 0 & 0 & 0 \\
0 & 0 & -\lambda & 0 & 0 & 0 \\
0 & 0 & -\lambda & 0 & 0 & 0 \\
0 & 0 & -\lambda & 0 & 0 & 0
\end{matrix}\right)\notag,
\end{align}

\begin{align}
A_{31}=\left(\begin{matrix}
0 & 0 & i\Lambda_{1} & 0 & 0 & 0 \\
0 & 0 & -i\Lambda_{1}^{\ast} & 0 & 0 & 0 \\
0 & \Lambda_{1} & 0 & 0 & 0 & 0 \\
\Lambda_{1}^{\ast} & 0 & 0 & 0 & 0 & 0 \\
0 & 0 & 0 & 0 & i\Lambda_{2} & 0 \\
0 & 0 & 0 & 0 & -i\Lambda_{2}^{\ast} & 0
\end{matrix}\right)\notag,
\end{align}

\begin{align}
A_{32}=\left(\begin{matrix}
0 & 0 & 0 & 0 & 0 & 0 \\
0 & 0 & 0 & 0 & 0 & 0 \\
\Lambda_{1}^{\ast} & 0 & i\Lambda_{1} & 0 & 0 & 0 \\
0 & \Lambda_{1} & -i\Lambda_{1}^{\ast} & 0 & 0 & 0 \\
0 & 0 & 0 & 0 & 0 & 0 \\
0 & 0 & 0 & 0 & 0 & 0
\end{matrix}\right)\notag,
\end{align}

\begin{align}
A_{33}=\left(\begin{matrix}
-\Omega_{1,+} & 0 & \omega_{m,1} & 0 & 0 & 0 \\
0 & \Omega_{1,-} & 0 & \omega_{m,1} & 0 & 0 \\
-\omega_{m,1} & 0 & -\Omega_{2,+} & 0 & -\lambda & 0 \\
0 & -\omega_{m,1} & 0 & \Omega_{2,-} & 0 & -\lambda \\
0 & 0 & 0 & 0 & -\Omega_{1,+} & 0 \\
0 & 0 & 0 & 0 & 0 & \Omega_{1,-}
\end{matrix}\right)\notag,
\end{align}

\begin{align}
A_{34}=\left(\begin{matrix}
0 & 0 & i\Lambda_{2} & 0 & 0 & 0 \\
0 & 0 & -i\Lambda_{2}^{\ast} & 0 & 0 & 0 \\
0 & 0 & 0 & 0 & 0 & i\Lambda_{2} \\
0 & 0 & 0 & 0 & 0 & -i\Lambda_{2}^{\ast} \\
\omega_{m,1} & 0 & i\Lambda_{1} & 0 & 0 & 0 \\
0 & \omega_{m,1} & -i\Lambda_{1}^{\ast} & 0 & 0 & 0
\end{matrix}\right)\notag,
\end{align}

\begin{align}
A_{41}=\left(\begin{matrix}
0 & \Lambda_{2} & 0 & 0 & 0 & 0 \\
\Lambda_{2}^{\ast} & 0 & 0 & 0 & 0 & 0 \\
0 & 0 & 0 & 0 & 0 & 0 \\
0 & 0 & 0 & 0 & 0 & 0 \\
0 & 0 & -\lambda & 0 & 0 & 0 \\
0 & 0 & 0 & 0 &-\lambda & 0
\end{matrix}\right)\notag,
\end{align}

\begin{align}
A_{42}=\left(\begin{matrix}
\Lambda_{2}^{\ast} & 0 & 0 & 0 & i\Lambda_{2} & 0 \\
0 & \Lambda_{2} & 0 & 0 & -i\Lambda_{2}^{\ast} & 0 \\
0 & 0 & 0 & 0 & 0 & 0 \\
0 & 0 & 0 & -\lambda & -\lambda & 0 \\
0 & 0 & 0 & 0 & 0 & 0 \\
0 & 0 & 0 & 0 & 0 & 0
\end{matrix}\right)\notag,
\end{align}

\begin{align}
A_{43}=\left(\begin{matrix}
-\lambda & 0 & 0 & 0 & -\omega_{m,2} & 0 \\
0 & -\lambda & 0 & 0 & 0 & -\omega_{m,2} \\
0 & 0 & 0 & 0 & 0 & 0 \\
0 & 0 & \Lambda_{2}^{\ast} & \Lambda_{2} & 0 & 0 \\
\Lambda_{2}^{\ast} & \Lambda_{2} & 0 & 0 & 0 & 0 \\
0 & 0 & 0 & 0 & \Lambda_{1}^{\ast} & \Lambda_{1}
\end{matrix}\right)\notag,
\end{align}

\begin{align}
A_{44}=\left(\begin{matrix}
-\Omega_{3,+} & 0 & 0 & 0 & i\Lambda_{1} & 0 \\
0 & \Omega_{3,-} & 0 & 0 & -i\Lambda_{1}^{\ast} & 0 \\
0 & 0 & 0 & 0 & \omega_{m,2} & \omega_{m,1}\\
\Lambda_{1}^{\ast} & \Lambda_{1} & 0 & -\Omega_{4} & -\omega_{m,1} & -\omega_{m,2} \\
0 & 0 & -\omega_{m,2} & \omega_{m,1} & -\gamma_{m,2} & 0 \\
0 & 0 & -\omega_{m,1} & \omega_{m,2} & 0 & -\gamma_{m,1}
\end{matrix}\right)\notag.
\end{align}

%

\end{document}